\documentclass[letterpaper,twocolumn,10pt]{article}
\usepackage{usenix2019_v3}
\usepackage{xurl}
\usepackage[scaled=0.84]{beramono}  %
\usepackage{amsmath}
\usepackage{amssymb}
\usepackage{enumitem}
\usepackage[ruled,vlined,linesnumbered]{algorithm2e}
\SetKwProg{Fn}{Function}{:}{}
\SetKwFunction{Attack}{Attack}
\SetKwFunction{Patch}{Patch}
\SetKwFunction{Converge}{Converge}
\newcommand{\Exec}{\text{\textsc{Exec}}}
\newcommand{\repro}{\text{\textsc{Repro}}}
\newcommand{\blocked}{\text{\textsc{Blocked}}}
\newcommand{\incon}{\text{\textsc{Incon}}}
\newcommand{\blk}[1]{\text{blocked}(#1)}
\newcommand{\noreg}{\text{noRegression}}
\newcommand{\katt}{\kappa_{\mathrm{att}}}
\newcommand{\kgate}{\kappa_{\mathrm{gate}}}

\newcommand{\headrow}{\rowcolor{p2gray}}

\usepackage{booktabs}
\usepackage{colortbl}
\usepackage{multirow}
\usepackage{tabularx}
\usepackage{xcolor}
\usepackage{hyperref}
\hypersetup{
  colorlinks=true,
  linkcolor=p2linkblue
}

\definecolor{p2blue}{RGB}{36,83,124}
\definecolor{p2blueLight}{RGB}{236,244,250}
\definecolor{p2green}{RGB}{45,104,77}
\definecolor{p2greenLight}{RGB}{235,246,240}
\definecolor{p2amber}{RGB}{150,95,26}
\definecolor{p2amberLight}{RGB}{252,244,225}
\definecolor{p2red}{RGB}{132,53,53}
\definecolor{p2redLight}{RGB}{248,235,235}
\definecolor{p2gray}{RGB}{246,247,249}
\definecolor{p2linkblue}{RGB}{36,83,124} %

\newcolumntype{L}{>{\raggedright\arraybackslash}X}
\usepackage{graphicx}
\graphicspath{{./}{figures/}}
\newif\ifgenfig
\genfigtrue
\usepackage{xspace}
\usepackage{tikz}
\usetikzlibrary{arrows.meta,fit,positioning,shapes.geometric,calc,backgrounds,decorations.pathreplacing}

\newcommand{\sys}{\textsc{P2Patch}\xspace}
\newcommand{\pov}{PoV\xspace}
\newcommand{\povs}{PoVs\xspace}
\newcommand{\code}[1]{\texttt{#1}}
\newcommand{\Cfix}{C_{\mathrm{fix}}}
\newcommand{\Cres}{C_{\mathrm{res}}}
\newcommand{\Sfix}{S_{\mathrm{fix}}}
\newcommand{\Sres}{S_{\mathrm{res}}}
\newcommand{\cc}{Claude Code\xspace}
\newcommand{\base}{\textsc{SinglePass}\xspace}
\newcommand{\bench}{\textsc{VulnRepairBench}\xspace}

\newcommand{\legacyparagraph}[1]{\vspace{1mm}\noindent\textbf{#1}}
\newcommand{\paperparagraph}[1]{\paragraph{#1}}

\newcommand{\hlg}[1]{{\setlength{\fboxsep}{0pt}\colorbox{p2green!30}{\strut #1}}}

\newcommand{\best}[1]{\textcolor{p2blue}{\textbf{#1}}}
\newcommand{\gain}[1]{\textcolor{p2green}{#1}}

\newcommand{\takeaway}[1]{%
  \par\smallskip\noindent
  \fcolorbox{black!20}{p2gray}{%
    \parbox{\dimexpr\columnwidth-2\fboxsep-2\fboxrule\relax}{%
      \small\textbf{Takeaway.} #1}}%
  \par\smallskip}

\newcommand{\full}{\textcolor{p2green}{$\checkmark$}}    %
\newcommand{\partialblk}{\textcolor{p2amber}{$\triangle$}} %
\newcommand{\zeroblk}{\textcolor{p2red}{$\times$}}       %
\newcommand{\nopatch}{\textcolor{black!45}{$\varnothing$}}      %
\newcommand{\notmeas}{\textcolor{black!45}{\textsf{?}}}        %
\newcommand{\na}{\textemdash}                                   %

\definecolor{csink}{RGB}{248,235,235}   %
\definecolor{cfix1}{RGB}{235,246,240}   %
\definecolor{cbyp}{RGB}{252,244,225}    %
\definecolor{cfix2}{RGB}{236,244,250}   %
\definecolor{cup}{RGB}{242,239,249}     %
\definecolor{cfix2g}{RGB}{234,245,238}  %

\newsavebox{\panelbuf}
\newlength{\panelpad}
\newcommand{\paneltag}[2]{{\sffamily\fontsize{6.8}{8}\selectfont\bfseries
  \textcolor{#1}{\MakeUppercase{#2}}\par\vspace{3pt}}}
\newenvironment{panelbox}[3]{%
  \def\panelfill{#1}\def\panelaccent{#2}%
  \par\addvspace{6pt}%
  \begin{lrbox}{\panelbuf}%
  \begin{minipage}{\dimexpr\columnwidth-2\panelpad-2\fboxrule\relax}%
    \setlength{\parindent}{0pt}\footnotesize
    \paneltag{#2}{#3}%
}{%
  \end{minipage}%
  \end{lrbox}%
  {\setlength{\fboxsep}{\panelpad}%
   \noindent\fcolorbox{\panelaccent!40}{\panelfill}{\usebox{\panelbuf}}}%
  \par\addvspace{7pt}}
\newenvironment{codepanel}[3]{%
  \panelbox{#1}{#2}{#3}\ttfamily\color{black!88}\raggedright\linespread{1.06}\selectfont
}{%
  \endpanelbox}
\newcommand{\ci}{\hspace*{1.4em}}
\newcommand{\hlp}[1]{{\setlength{\fboxsep}{0pt}\colorbox{p2amber!30}{\strut #1}}}


\begin{document}
\date{}
\title{\Large\bf \sys: Exploit-Guided Vulnerability Repair}
\author{
{\rm Mohsen Iranmanesh$^{1*}$, Sina Moradi Sabet$^{2*}$, Sina Marefat$^{3}$, Ali Javidi Ghasr$^{4}$,}\\
{\rm Maxime Newman Nereyabagabo$^{1}$, Kousha Amouzesh$^{1}$, Jaivir Malhi$^{1}$,}\\
{\rm Vinci Velasco$^{1}$, Alex Chung$^{1}$, Mohammad A. Tayebi$^{1}$}\\[6pt]
{\normalsize\rm $^{1}$Simon Fraser University, Canada \quad $^{2}$Amirkabir University of Technology, Iran}\\
{\normalsize\rm $^{3}$K. N. Toosi University of Technology, Iran \quad $^{4}$Ferdowsi University of Mashhad, Iran}\\[4pt]
{\small\rm mia32@sfu.ca, sina.moradi@aut.ac.ir, sina.marefat@email.kntu.ac.ir, alijavidighasr@mail.um.ac.ir,}\\
{\small\rm mnn6@sfu.ca, kousha\_amouzesh@sfu.ca, jai\_malhi@sfu.ca, vvv@sfu.ca, alex\_chung@sfu.ca, tayebi@sfu.ca}
}
\maketitle
{\renewcommand{\thefootnote}{\fnsymbol{footnote}}\footnotetext[1]{Equal contribution.}}

\begin{abstract}
Large language models are increasingly deployed as autonomous agents for automated program repair, patching software vulnerabilities through multi-step reasoning and tool use. Yet these systems typically validate a patch against the same evidence used to construct it---a static alert, sanitizer trace, published exploit, or test suite. A patch that blocks that one piece of evidence can therefore look successful while remaining vulnerable to the same vulnerability with some variation. We present \sys, a generative adversarial repair system that treats each patch as a new attack target. Starting from a normalized vulnerability finding, an exploitation agent constructs a reproducing proof of vulnerability (PoV), a patching agent repairs the program, and the exploitation agent then attacks the patched program for a distinct same-vulnerability exploit. Every successful exploit becomes a permanent repair requirement, and the final patch must block all PoVs accumulated during patching process, with construction-time PoVs kept strictly separate from held-out evaluation oracles. Across 101 C/C++ and Java CVE subjects, this adversarial loop surfaces a distinct confirmed exploit in 41\% of runs after the patch has already blocked its original construction-time PoV.
Against single-pass repair from the same finding using Claude Code, \sys raises held-out fix-closed coverage from 78.1\% to 84.7\% and beyond-upstream coverage from 42.1\% to 62.5\%. Under the same held-out oracles, it also certifies more attack paths as blocked than recent state-of-the-art agentic repair systems on overlapping subjects. These results show that adversarial testing and independent security oracles expose repair gaps that one-shot validation misses.
\end{abstract}
\section{Introduction}

Automated vulnerability repair is only as reliable as the evidence used to
guide and judge it. A static-analysis alert identifies a suspicious flow, but
not necessarily the runtime conditions that make it exploitable~\cite{johnson2013static,kang2022falsealarms}. A sanitizer
trace or proof of concept---an input that triggers a specific bug or flaw within a program---demonstrates one failing execution, but not every similar payload, encoding, entry point, or sibling path that should be blocked.
A project test suite may preserve intended functionality, but it rarely defines
the full security boundary. These signals are valuable but partial: they can
explain why a revision merits repair without specifying what a complete repair
must cover~\cite{hu2025sok,li2025effectivesok}.

These partial signals create an oracle-reuse problem. Automated repair systems often
use one piece of evidence twice: first to guide patch construction, and then to
decide whether the patch succeeded. A patch that blocks the supplied \pov,
crashing input, sanitizer trace, or available test may therefore be accepted
even if it only specializes to that evidence. This risk is well known in
general program repair, where test-guided repair can overfit its construction
suite~\cite{smith2015overfitting,yang2017bettertests,le2018semanticoverfitting,legoues2012genprog}. It is especially consequential for
security repair: blocking one observed attack can be mistaken for resolving the
underlying security issue. Recent vulnerability-repair evaluations likewise
show that proof-of-concept plus functional-test validation can overestimate
repair correctness~\cite{yu2026pvbench,wang2025vulnrepaireval,wei2025patcheval}.

Existing vulnerability-repair systems occupy several points in this design
space. Some patch from vulnerable code, weakness semantics, and suspected
locations, while others use crashing inputs, sanitizer reports, proofs of
concept, or trace-derived triggers for localization, repair, and
validation~\cite{nong2025appatch,gao2021extractfix,zhang2022vulnfix,yu2025patchagent,kim2025san2patch,zhang2026ossfuzz,joos2026codecureagent,shoshitaishvili2018mechphish,song2015cgc}.
LoopRepair further iterates on vulnerable locations and taint-trace
feedback~\cite{ye2026looprepair}. These systems show that dynamic evidence and
LLM agents can improve vulnerability repair, but their repair objective is
typically fixed before patching begins: the system attempts to satisfy the same
alert, crash, trace, or \pov after each repair attempt. The target does not
adapt to the patched program.

We argue that vulnerability repair should be
\emph{exploit-guided}. Once a patch changes the program, the relevant
question is not only whether the original \pov is blocked, but whether a
distinct exploit can falsify the security claim that accepting the current patch
implies. A path-traversal patch that rejects one relative path may
still accept an absolute path; an injection patch that blocks one payload may
miss an alternate encoding; a repair at one sink may leave a sibling sink
reachable. In these cases, the next useful evidence is not another explanation
of the original bug, but a working same-vulnerability exploit against the current patch.
This framing resembles counterexample-guided synthesis, but our verifier is an
incomplete, budgeted exploiter rather than a complete decision
procedure~\cite{solar2006combinatorial}.

This paper presents \sys, an execution-refereed exploit--patch repair system. We call this \emph{exploit-guided} repair: a security instantiation of counterexample-guided synthesis in which the counterexample is an executable exploit against the current patch rather than a logical model~\cite{solar2006combinatorial,clarke2000cegar}. \sys starts from a normalized vulnerability finding and a pinned project revision. The protocol is evidence-source independent: the finding may come from a static alert or source-to-sink trace~\cite{arzt2014flowdroid}, a vulnerability description, a sanitizer report, or a successful exploitation. Our evaluation uses source-to-sink traces, with all findings redacted so that agents do not receive the CVE identifier, upstream fix, advisory exploit text, or held-out proofs of vulnerability (\povs). Existing work has studied LLM-assisted static analysis and \pov generation as upstream capabilities~\cite{li2025iris,nitin2025faultline,gezgin2026pocgym,ullah2025cvegenie,li2024llift}. \sys uses such evidence differently: an exploitation agent (the \emph{exploiter}) turns the finding into a reproducing \pov, a patching agent (the \emph{patcher}) modifies the program until that \pov is blocked, and the exploiter then attacks the patched program for a new distinct same-vulnerability \pov that bypasses the patch. Every execution-confirmed bypassing \pov becomes a permanent repair requirement.

The design is execution-mediated. Agents may propose \povs and patches, but
they do not certify success~\cite{zheng2023judge}. The orchestrator executes \povs in the project
container, distinguishes reproduction, blocking, and inconclusive
infrastructure outcomes, runs differential regression checks, protects
role-specific artifacts, and replays accumulated hardening \povs after the
adversarial rounds. Thus, the interaction is a bounded repair process over a changing
executable program, not a natural-language debate between agents.

\sys separates construction-time evidence from evaluation-time evidence.
Live \povs drive repair, but we do not treat them as an independent measure
of completeness to avoid the oracle-reuse problem. We instead evaluate final
patches with independently certified and held-out PoV suites. Fix-closed PoVs
reproduce on the vulnerable revision and are blocked by the recorded official
fix, measuring held-out attack-path coverage. Residual PoVs reproduce both
before and after the official fix, measuring beyond-upstream hardening. This
distinction matters because released security patches can themselves be
incomplete~\cite{li2017securitypatches,yin2011incorrectfixes}.

We evaluate \sys on 101 project-level CVE subjects spanning C/C++ and Java. The exploiter finds a distinct execution-confirmed bypass in 41\% of runs after the patch blocks its original \pov. Against single-pass repair from the same finding, \sys raises held-out fix-closed coverage from 78.1\% to 84.7\% and beyond-upstream coverage from 42.1\% to 62.5\%. Fourteen certified residual gaps were later closed upstream, up to nine and a half years after the original fix. Under the same held-out oracles, \sys covers more certified attack paths than published repair systems on overlapping subjects and exposes patches that passed their validation while leaving certified paths open.

This paper makes the following contributions:
\begin{itemize}[leftmargin=*]
    \item We introduce a counterexample-guided exploit--patch protocol in which
    each execution-confirmed bypass of the current patch becomes a permanent
    repair requirement.

    \item We design an execution-mediated orchestrator that separates exploiter
    and patcher roles, protects role-specific artifacts, distinguishes blocked
    exploits from inconclusive outcomes, performs differential regression
    checks, and replays accumulated \povs.

    \item We release \bench, a benchmark of 101 project--CVE subjects whose
    independently certified fix-closed and residual \pov suites separate
    held-out attack-path coverage from beyond-upstream hardening.

    \item We evaluate \sys across C/C++ and Java CVE subjects against a
    single-pass baseline, loop ablations, model variants, and published repair
    systems, showing that adaptive counterexample growth improves security
    coverage and exposes gaps missed by conventional validation.
\end{itemize}

\begin{figure*}[t]
\centering
\includegraphics[width=\linewidth]{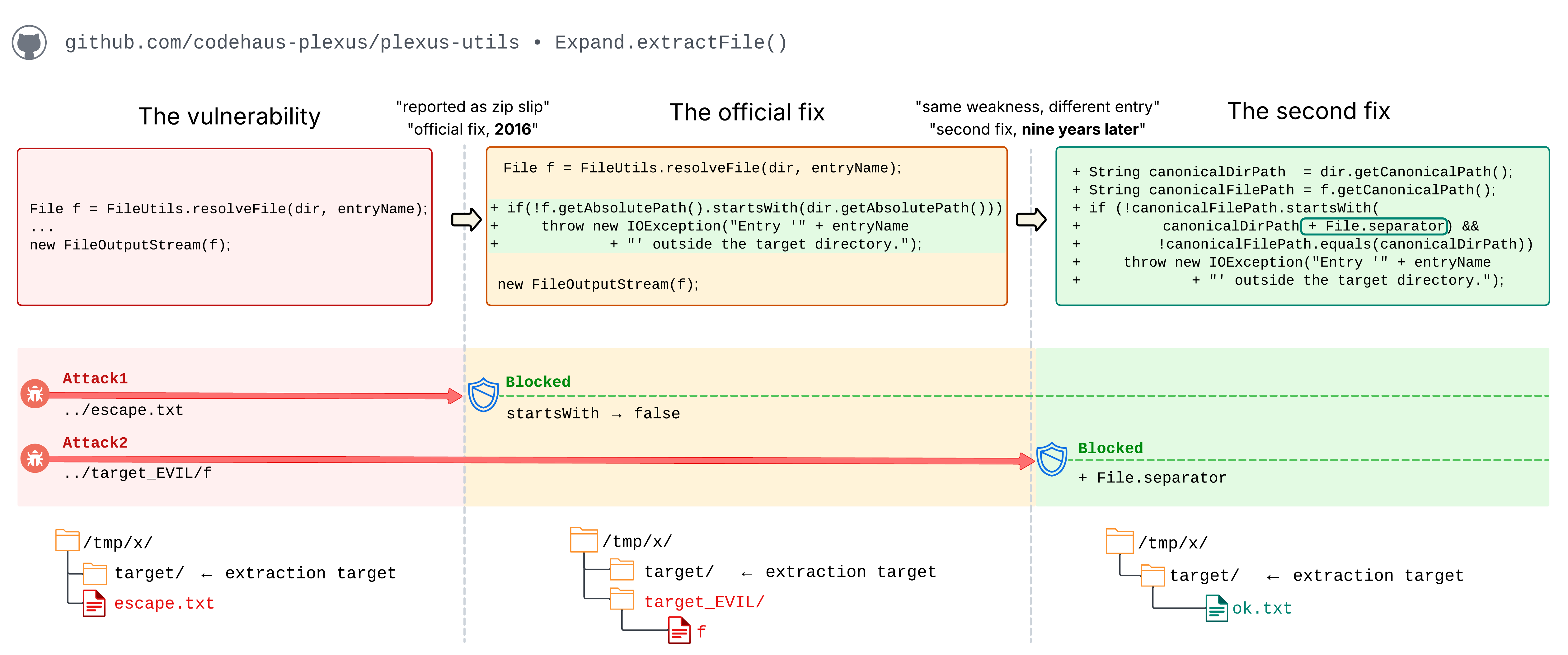}
\caption{Motivating \code{plexus-utils} example. The 2016 containment check is bypassed by a sibling path; the later fix closes the gap by canonicalizing paths and enforcing a separator boundary.}
\label{fig:motivation}
\end{figure*}

\section{Motivation and Problem Setting}
\label{sec:problem}

\sys studies repair after a vulnerability has already been reported. Given a
pinned vulnerable revision $P_0$, a normalized finding $f$, and a project
execution contract $\mathcal{C}$, the task is to produce either a source patch
$\pi$ or a structured failure. A patch is construction-valid only if the
patched program $P_0\oplus\pi$ (the pinned revision with $\pi$ applied) blocks
all live \povs accumulated during the protocol and
does not introduce a differential regression under $\mathcal{C}$. Final
security coverage is measured later by held-out \pov suites; \sys is blind to these during repair.

We study repair, not triage~\cite{heckman2011actionable}. Each subject corresponds to a reported CVE, and
the normalized finding is assumed to describe the target vulnerability. The
system receives the weakness class, relevant source locations, suspected
source-to-sink or trigger context when available, and build/execution metadata.
It does not receive the CVE identifier, upstream patch, advisory exploit text,
or curated evaluation \povs. This setup evaluates whether a repair protocol can
turn partial vulnerability evidence into executable repair requirements, not
whether an agent can retrieve or copy the recorded fix.

\subsection{One Vulnerability, Two Fixes}

Figure~\ref{fig:motivation} gives the motivating example. plexus-utils, a
utility library used throughout the Maven build ecosystem, extracts archives
through \code{Expand.extractFile}. It resolved an entry name against the
extraction directory and opened the result for writing, with no containment
check in between:

\begin{codepanel}{csink}{p2red}{The vulnerability}
File f = FileUtils.resolveFile(dir, entryName);\\
\ldots\\
new FileOutputStream(f);
\end{codepanel}

An archive entry named \code{../escape.txt} therefore resolves above the
extraction directory and is written there---the archive-extraction path
traversal known as zip slip~\cite{snyk2018zipslip}. This is the reported vulnerability, and the
project's 2016 fix inserts a containment check immediately after the entry is
resolved: reject the entry unless its absolute path \code{startsWith} the
extraction directory's absolute path. Against the reported entry the fix
works, and it works against every deeper \code{../} chain that lands above the
target, since each of those resolves somewhere that does not begin with the
target's path at all.

The check is nevertheless incomplete, and the reason is one line of it.
\code{String.startsWith} compares characters, not filesystem path components, and
\code{dir} contributes no trailing separator. An entry that resolves into a
\emph{sibling} of the extraction directory---one whose name merely extends the
target's---therefore walks straight through the guard:

\begin{panelbox}{cbyp}{p2amber}{The bypass --- a sibling, not a descendant}
\begin{tabular}{@{}l@{\hspace{7pt}}l@{}}
\ttfamily\hlp{/tmp/x/target} & the extraction target\\[2pt]
\ttfamily\hlp{/tmp/x/target}\_EVIL/f & where \code{../target\_EVIL/f} lands\\
\end{tabular}
\par\medskip
The shaded characters are identical, so the comparison returns \textbf{true}.
The guard stays silent and the entry is written outside the very directory the
guard exists to protect.
\end{panelbox}

\noindent The patched program remains vulnerable to the weakness it was
patched for, through a directory name that merely extends the one it protects.
More than nine years later, upstream replaced the check with one that
canonicalizes both operands and compares against the target path \emph{plus a
separator}:

\begin{codepanel}{cfix2g}{p2green}{The second fix, 2025}
if (!canonicalFilePath.startsWith(\\
\ci canonicalDirPath \hlg{ + File.separator})\\
\ci \&\& !canonicalFilePath.equals(canonicalDirPath))
\end{codepanel}

\noindent The separator moves the comparison onto a path boundary:
\code{/tmp/x/target/} is not a prefix of \code{/tmp/x/target\_EVIL/f}, so the
sibling is rejected while real descendants still pass. The second fix does not
correct a misunderstanding about the weakness or the sink; the first fix had
both right. It corrects the boundary. The same boundary-less check appears in
other projects' official fixes in our benchmark; Appendix~\ref{app:upstream}
measures how long those gaps survived.

The example shows a gap between two questions a repair system can
confuse: whether a patch stops the attack it was shown, and whether it repairs
the boundary that attack crossed. Re-running the reported entry answers the
first. Nothing used to construct the patch answers the second---closing that gap
requires evidence independent of the repair process.

\subsection{From Findings to Repair Requirements}

The evidence available for a vulnerability determines what a repair system can
be asked to do. A sanitizer report plus a crashing input is precise and
executable, but it exists only for failures a sanitizer or crash oracle can
observe. A static finding is broader, covering injection, traversal,
authorization, deserialization, and other non-crashing weaknesses. On its own,
though, it neither proves reachability nor defines the observable security effect~\cite{yamaguchi2014cpg}.
\sys consumes the latter form and turns it into live \povs.

A \emph{proof of vulnerability} (\pov) is an input or test driver plus an
execution command whose outcome is one of three values: reproduced, blocked, or
inconclusive. A \pov is \emph{live} when the exploiter produces it
during a run, so that it becomes a repair requirement; it is \emph{held-out}
when it is authored outside the loop and used only to score the final patch. A
valid \pov must exercise a real product entry point and observe the relevant
security effect. Source inspection, attacker-controlled assertions, missing
harnesses, timeouts, and build failures are not evidence that a vulnerability is
blocked. The inconclusive outcome is therefore essential: without it, broken
infrastructure can be mistaken for successful repair.

A \pov is both narrower and stronger than the original finding. It is narrower
because it demonstrates one path, payload, and environment. It is stronger
because it converts a suspected vulnerability into an executable predicate that
can be replayed before and after patching. The repair objective starts from such
predicates and grows only when execution confirms another in-scope bypass.

\subsection{Growing Objective and Scope}

The repair objective is not fixed to the first \pov. When execution confirms a
new bypass of the current patch, that bypass becomes an additional repair
requirement that later patches must continue to block. New requirements must
exercise the same reported weakness and security boundary, although they may
use different payloads, encodings, entry points, or sibling paths.

This structure resembles counterexample-guided synthesis, but the analogy is
methodological rather than formal~\cite{alur2013sygus}. The verifier is a bounded exploiter,
not a complete decision procedure over all inputs~\cite{cadar2008klee,jha2010oracle}. ``No counterexample found''
therefore means only that the configured search failed to find another
reproducing same-vulnerability bypass. Independent held-out suites in
Section~\ref{sec:oracles} measure the final patch after repair terminates.

\subsection{Threat Model and Non-Goals}

\sys targets known vulnerabilities in pinned C/C++ and Java project revisions.
The attacker controls inputs relevant to the finding, such as archive entries,
files, buffers, request parameters, serialized objects, command-line arguments,
or network messages. We assume the revision can be built and exercised in its
container and that the vulnerable effect is observable under the project
contract. Regression checks are bounded behavioral oracles: passing them
establishes compatibility only for the behaviors they exercise.

The agents receive the normalized finding, source tree, and build metadata, but not the CVE identifier, upstream fix, advisory exploit text, or held-out suites, and web retrieval is disabled. These controls reduce direct leakage, but they do not remove knowledge acquired during pretraining. Each run uses a dedicated Git worktree and project container. The prototype does not claim host containment against a malicious repository. We treat the evaluation host and benchmark repositories as trusted infrastructure, and agent-produced edits, commands, and claims as untrusted inputs.

\section{Counterexample-Guided Repair Protocol}
\label{sec:design}

\begin{figure*}[t]
    \centering
    \makebox[\textwidth][c]{%
        \includegraphics[width=1\textwidth, keepaspectratio]{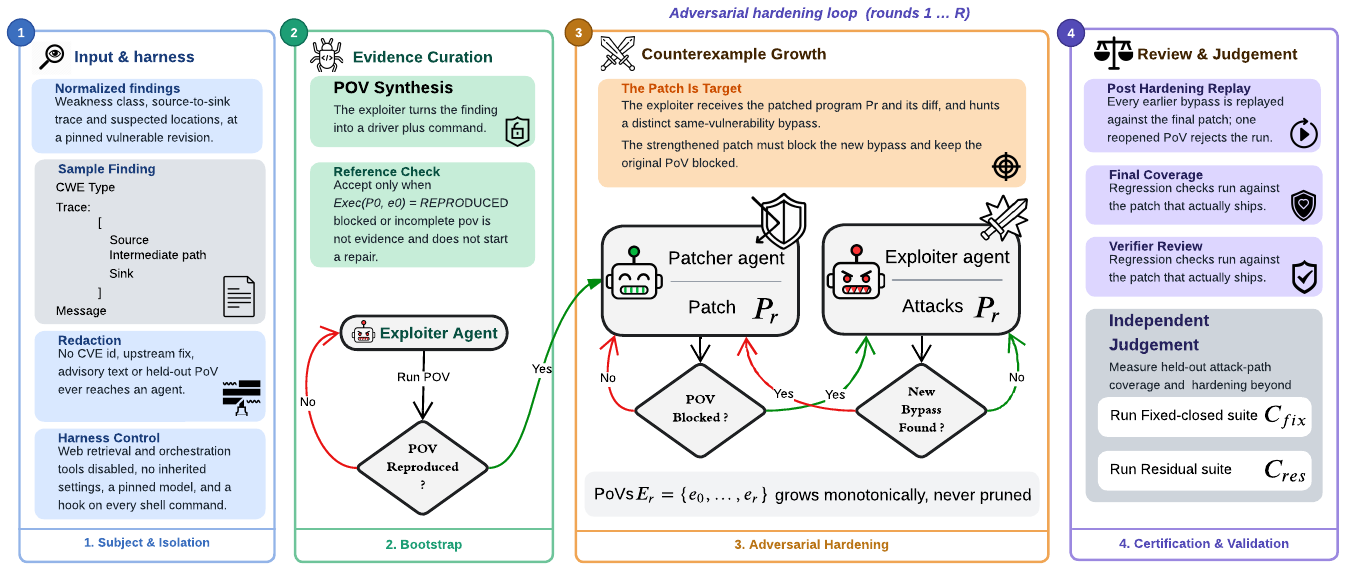}%
    }
    \caption{\sys workflow: bootstrap a live \pov, patch accumulated requirements,
    attack the patched program for bypasses, then score the final patch with
    held-out suites.}
    \label{fig:overview}
\end{figure*}

\sys turns the problem setting above into an execution-refereed repair protocol
(Figure~\ref{fig:overview}). The protocol is intentionally simple: make the
reported finding executable, patch the current requirement set, attack the
patched program for a same-vulnerability bypass, and repeat only when execution
confirms a new bypass. The full pseudocode is in
Appendix~\ref{app:protocol} and the agent prompt templates ship with the
artifact (Appendix~\ref{sec:openscience}); this section describes the design
invariants and runtime semantics needed to understand the evaluation.

\subsection{Protocol State and Invariants}

A run maintains three pieces of state: the pinned vulnerable revision $P_0$,
the current candidate program $P_r$, and an append-only requirement set
\[
E_r = \{e_0,\ldots,e_r\}.
\]
Each $e_i \in E_r$ is an execution-confirmed live \pov that remains a repair
obligation. Initially, $E_{-1}=\emptyset$. During adversarial hardening,
whenever execution confirms a new same-vulnerability bypass $e_{r+1}$, the
orchestrator extends the requirement set as
\[
E_{r+1}=E_r\cup\{e_{r+1}\}.
\]

We write $b=|E|-1$ for the number of execution-confirmed bypasses a run appends
beyond the bootstrap \pov $e_0$; Tables~\ref{tab:persubject}
and~\ref{tab:persubjectc} report $b$ per subject.

This state enforces three invariants. First, requirements are
\emph{executable}: a finding becomes a repair requirement only after its \pov
reproduces in the project container. Second, requirements are
\emph{persistent}: once a \pov enters $E_r$, it remains a repair obligation
for the rest of the run. Third, agents do not certify their own success:
the orchestrator decides \pov reproduction, patch blocking, regression
outcomes, and final replay through execution.

The requirement set is a construction-time object. Satisfying every live
requirement does not establish patch completeness; final security coverage is
measured separately using the held-out oracle suites of
Section~\ref{sec:oracles}.

\subsection{Bootstrap: Making the Finding Executable}

Bootstrap is the bridge from partial vulnerability evidence to an executable
repair task. The exploiter receives the normalized finding, source tree, and
project execution contract, and tries to produce a \pov $e_0$ that
demonstrates the reported behavior on $P_0$. This \pov is not an evaluation
oracle. It is construction evidence: it establishes reachability, fixes the
runtime environment, and creates a concrete predicate that can be replayed while
the patcher edits code.

A candidate \pov is accepted only when \textsc{Exec}$(P_0,e_0)$ returns
\textsc{Repro}. If the candidate times out, fails to build, depends on a missing
harness, or does not observe the vulnerable behavior, it is inconclusive or
blocked and does not enter the requirement set. If no reproducing \pov is
found within the bootstrap budget, \sys returns a structured failure rather
than asking the patcher to repair an unexecuted claim. This conservative rule is
important: a non-reproducing \pov can make a broken harness look like a
successful patch, and a patch derived from an unexecuted claim cannot be scored
as construction-valid.

Once $e_0$ is accepted, it grounds the reported vulnerability in a reproducible
execution. The repair objective is to address the vulnerability while preserving
the project's intended behavior, not merely to block this \pov. The
accepted \pov initializes $E_0=\{e_0\}$, and the patcher receives the current
source tree, the normalized finding, the \pov, and execution feedback from the
orchestrator. The CVE identifier, upstream patch, advisory exploit text, and
held-out evaluation \povs remain unavailable.

\subsection{Patching Current Requirements}

Given a requirement set $E_r$, the patcher edits product code and proposes a
candidate program $P'_r$. The orchestrator then applies a security-first gate.
Regression checks are considered only after every \pov the gate protects
executes as \textsc{Blocked}. The first patch is gated on $e_0$; each hardening
round is gated on that round's new bypass together with $e_0$; earlier confirmed
bypasses are replayed once after the hardening loop, and the differential
regression check runs afterward.
If an earlier bypass fails this replay, the
patch is rejected without another correction attempt. If a \pov protected by a
convergence gate still reproduces, the patcher receives execution feedback for
another correction attempt. If the patch breaks behavior that passed on the
comparison revision, it is also rejected. The inner loop continues until the
gate passes or the correction budget is exhausted. Patches that pass the automated gates are subsequently subjected to the
standardized manual-review protocol of Section~\ref{sec:manual-review}.

The per-round gate is deliberately stronger than checking the most recent \pov:
it checks the newest bypass together with $e_0$. Because $E_r$ is append-only,
earlier bypasses remain repair obligations and are replayed after hardening. This
prevents a candidate from blocking $e_r$ by reopening $e_0$, while the
post-hardening replay rejects a candidate that reopens another earlier \pov. The
predicate sequence of each convergence gate is re-evaluated from the beginning
after every revision, so a regression fix cannot silently reintroduce $e_0$.

Patch convergence operates over the current requirement set; adversarial
hardening instead expands that set by searching the accepted patch for a new
same-vulnerability bypass. Section \ref{sec:hardening} describes this outer loop.

\subsection{Adversarial Hardening}
\label{sec:hardening}
After a candidate patch passes the current gate, \sys gives the patched program
to the exploiter. The exploiter's task is no longer to reproduce the vulnerable
revision. It must find a distinct same-vulnerability bypass against the current
patch. In other words, the patch becomes the next attack target. If the
exploiter produces a candidate $e_{r+1}$ and execution confirms
\textsc{Exec}$(P_r,e_{r+1})=\textsc{Repro}$, the candidate is appended to the
requirement set, $E_{r+1}=E_r\cup\{e_{r+1}\}$, and the patcher must strengthen
the repair against the expanded set.

The outer loop is bounded. The main configuration uses a fixed maximum number
of hardening rounds, and a stable exit means only that no additional confirmed
bypass was found before the budget expired. The final patch is therefore a
bounded hardening result, not a proof of vulnerability absence. Its security is
measured later by held-out oracle suites that were not used to construct the
patch.

\subsection{Execution Referee}
\label{sec:referee}

The orchestrator, not either agent, is the referee. For a program $P$ and
\pov $e$,
\[
\textsc{Exec}(P,e)\in\{\textsc{Repro},\textsc{Blocked},\textsc{Incon}\}.
\]
\textsc{Repro} means the \pov executes and observes the vulnerable behavior.
\textsc{Blocked} means the \pov executes but the vulnerable behavior is not
observed under the \pov exit contract. \textsc{Incon} means execution does not
establish either result because of a timeout, build failure, missing harness,
harness error, or other infrastructure failure.

The three-valued outcome avoids unsafe shortcuts. Bootstrap accepts a \pov
only on \textsc{Repro}; patch convergence accepts a transition only when all
protected \povs are \textsc{Blocked} and regression checks pass.
\textsc{Incon} helps neither agent. Treating inconclusive execution as
``blocked'' would reward deleted harnesses, stale builds, and infrastructure
failures. Treating it as ``reproduced'' would turn tool failures into false
security findings. \sys therefore gives no security credit for inconclusive
runs.

Regression checks are differential. A nonzero test result after patching counts
against the patch only when a fresh comparison against the pristine revision
supports that attribution. This matters because injected \pov files
may be discovered by project test runners, projects may have flaky or
environment-specific failures, and a stale build can fail for reasons unrelated
to the proposed repair. The same execution semantics are used for live gates and
final replay, while held-out evaluation suites are mounted only after repair
terminates.

\subsection{Artifact Integrity and Implementation}
\label{sec:impl}

The exploiter and patcher operate on the same changing program, but their
permissions are asymmetric. The exploiter may create attack artifacts but must
not weaken product code. The patcher may edit product code but must not delete,
rewrite, or neuter protected \povs. The orchestrator enforces this
separation with two role-specific guards: before exploiter calls, it snapshots
product sources and restores edits outside the \pov tree; before patcher calls,
it snapshots the \pov tree and restores edits, deletions, additions, or symlink
changes to protected \povs.

These guards are not a cosmetic implementation detail. Without them, the
exploiter could manufacture an easy bypass by weakening the program, and the
patcher could pass the gate by deleting or changing \povs instead of fixing
code. The guards make the requirement set meaningful: each \pov remains a
protected executable requirement across patch revisions, and each patch is judged
on product-code behavior rather than artifact tampering.

The prototype is a Python orchestrator whose configurations are typed stage
lists. Each stage declares required and produced artifacts, so configurations
are dependency-checked before execution. Build metadata records
project-specific toolchains and commands as ecosystem adapters rather than
language-specific repair logic. Each run persists agent inputs and outputs,
container logs, retry histories, diffs, and verdict files. The runner disables
web retrieval and inherited agent settings so that experimental arms change
repair evidence and loop structure, not the surrounding machinery. Full
pseudocode and implementation details are in
Appendix~\ref{app:protocol}.

\section{Independent Security Oracles}
\label{sec:oracles}
\begin{figure}[t]
\centering
\includegraphics[width=\columnwidth]{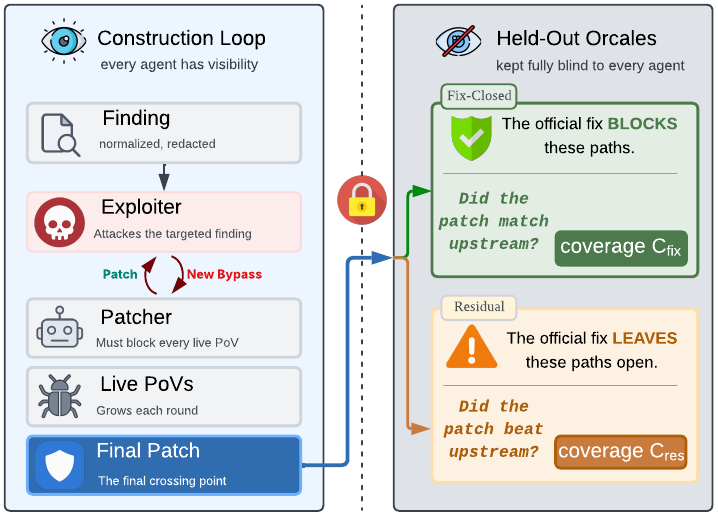}
\caption{Construction and held-out evaluation are strictly separated. During
repair, agents see only the normalized finding and accumulated live \povs.
After repair terminates, only the final patch crosses into held-out evaluation:
fix-closed oracles measure whether it blocks paths closed by the official fix
($\Cfix$), while residual oracles measure whether it blocks paths the official
fix leaves open ($\Cres$).}
\label{fig:oracle-families}
\end{figure}

Tests used during repair are insufficient as independent evidence of patch correctness~\cite{ye2021apcascale}. Similarly, live \povs are necessary for construction but cannot independently measure final security because the patcher has seen them. \sys therefore scores every patch with two held-out oracle families whose certification rules are opposites, while keeping both separate from the requirement set that drives repair (Figure~\ref{fig:oracle-families}).
Evidence used to construct the patch is never reused as the final measure of
security coverage. This separation is the central methodological constraint in
the paper: a patch may pass every live gate and still fail held-out paths that
were not available during repair.

The held-out suites are manually authored outside the repair loop, finalized
before scoring, and never shown to agents. They follow the same three-valued
execution contract as live \povs: a \pov can reproduce the vulnerability,
be blocked by the program under test, or be inconclusive because execution does
not establish either result. Inconclusive outcomes are excluded from the
coverage denominator and provide no security credit. These checks establish behavior under the
project execution contract; they are not complete semantic proofs.

\subsection{Fix-Closed Oracles}

A \emph{fix-closed \pov} is a held-out \pov that reproduces on the buggy
revision $P_{\mathrm{vuln}}$ and is blocked by the officially patched revision
$P_{\mathrm{up}}$:
\[
\textsc{Exec}(P_{\mathrm{vuln}},e)=\textsc{Repro}
\quad\text{and}\quad
\textsc{Exec}(P_{\mathrm{up}},e)=\textsc{Blocked}.
\]
Fix-closed suites measure coverage of independently constructed attack paths
that the recorded fix closes. They are not given to the exploiter as starting
points and are not given to the patcher as tests. They are mounted only after a
configuration ships a final patch. If a subject has five fix-closed \povs and
a generated patch blocks all five, then its fix-closed coverage for that
subject is 100\%; if it blocks only the path its live \pov drives, the held-out
suite exposes that specialization.

This oracle type gives the official fix an operational role without treating
it as source text to imitate. The generated patch is not compared syntactically
with the upstream diff. It is evaluated behaviorally: does it block the same
independently certified paths that the official fix blocks? This matters for
LLM repair because two patches can look very different while providing similar
coverage, and two superficially similar patches can differ on untested paths~\cite{xiong2018patchsim}.

\subsection{Residual Oracles}

A \emph{residual \pov} is a held-out \pov that reproduces on both the buggy
revision and the official fixed revision:
\[
\textsc{Exec}(P_{\mathrm{vuln}},e)=\textsc{Repro}
\quad\text{and}\quad
\textsc{Exec}(P_{\mathrm{up}},e)=\textsc{Repro}.
\]
Residual suites measure candidate gaps that appear to survive the recorded fix.
Because the official fix does not block them, they are certified by an inverted
rule rather than the fix-closed one. We therefore treat them as manually curated candidate
gaps in the official patch, and report them separately as $\Cres$. Later upstream commits that close the same behavior
provide additional evidence that a residual path was security-relevant, but they
are not required for inclusion.

Reporting fix-closed and residual suites separately avoids overclaiming. A
patch can match official-patch behavior by covering fix-closed paths while
still leaving nearby behaviors open; conversely, a patch that blocks residual
paths may be stronger than the recorded fix on those paths.

\subsection{Coverage and Leakage Control}
\label{sec:leakage}
For both oracle types, coverage is computed only over conclusive executions.
The main metric, defined in Section~\ref{sec:setup}, counts the fraction of
held-out \povs blocked by the generated patch and reports inconclusive outcomes
separately. The same execution referee decides live \povs, final replay, fix-closed
suites, and residual suites. No configuration can therefore receive credit
under a different acceptance rule than the one used during construction.

Oracle leakage is controlled by separation in time and interface. Agents do not
receive CVE identifiers, upstream patches, advisory exploit text, or held-out
oracle files; web retrieval is disabled; and held-out suites are mounted only
after repair terminates. These controls cannot eliminate knowledge present in a
model's pretraining data, so our claim is not that agents are uninformed about
the world. The claim is narrower and operational: the repair loop cannot inspect
or execute the evaluation suites that determine the reported coverage.
\section{Experimental Setup}
\label{sec:setup}

Our evaluation tests whether counterexample-guided hardening improves security
coverage beyond one-shot repair from that finding. We vary only whether
the patcher sees no executable \pov, a bootstrap \pov, or a growing set of
execution-confirmed bypasses. All configurations share subjects,
normalized findings, execution contracts, model pinning, artifact guards,
regression policy, and held-out oracle suites.

\paperparagraph{Benchmark.}
We evaluate on \bench, the benchmark we assemble for this work
(Appendix~\ref{app:benchmark}).
A subject is a project--CVE instance at a pinned vulnerable revision, with a
source tree, normalized finding, container definition, and project-specific
build, PoV-execution, and regression commands~\cite{jimenez2024swebench}. C/C++ subjects are derived
from VulnLoc+~\cite{ye2026looprepair,shariffdeen2021crashrepair}, restricted to
those whose required contracts could be reproduced. Java
subjects are curated from GitHub Security Advisories and the NVD; each uses a
manually checked CodeQL-derived taint finding consistent with the advisory and
vulnerable revision~\cite{anwar2022nvdquality,avgustinov2016ql}. The leakage controls of Section ~\ref{sec:leakage} apply to every subject and configuration.
\bench contains 101 CVEs (60 Java and 41 C/C++) over 61 distinct
repositories (51 Java and 10 C/C++), and spans 20 weakness classes with no
overlap between the two ecosystems. Every CVE has a certified fix-closed suite
with 1--14 \povs; residual suites exist only where the official fix leaves a
demonstrated gap, with 1--5 \povs each. Table~\ref{tab:corpus} summarizes these
totals.

\paperparagraph{Fairness controls.}
All agents run through the same harness: \cc (v2.1.241), a coding agent with
file, shell, and search tools. Every invocation is restricted to a per-run Git
worktree, with web retrieval and orchestration tools disabled, no inherited
settings, a pinned model, and a pre-execution command hook. This includes
\sys itself, so the arms differ only in the evidence and loop structure the
orchestrator imposes on that harness, not in the agent driving it. Configurations
receive identical inputs apart from the experimental variable, use the same
applicable per-stage budgets, and are scored by identical held-out suites. The
main comparison gives the exploiter three attempts when an exploitation stage
is present, each convergence gate three evaluations (hence at most two corrective
patches), and the hardening loop four rounds. Deviations from these budgets that never became binding are reported with the
results.

\paperparagraph{Patch acceptance and manual review.}
\label{sec:manual-review}
Manual review is performed only after a candidate patch passes the automated execution and regression gates. Two reviewers independently assess each candidate using a fixed rubric covering unintended functional changes, removal or weakening of security-relevant logic, build-system modifications, deletion or circumvention of tests or harnesses, and introduction of new security risks. Reviewers are blinded to the model and experimental configuration and base their decisions only on the patch diff, relevant program context, and execution evidence. Disagreements are resolved through joint adjudication. The same procedure is applied across all configurations. Held-out security oracles are not available during this review and do not influence patch acceptance.

\paperparagraph{Internal configurations.}
We compare configurations expressible as parameter settings of
Algorithm~\ref{alg:protocol}.
\begin{itemize}[leftmargin=*]
    \item \textbf{\base.} The patcher repairs the normalized finding in one
    pass on the same \cc harness, with no \Attack{}, no live \pov, and no post-patch bypass search.
    This measures repair from static evidence alone.
    \item \textbf{PoV-only.} This configuration keeps the bootstrap
    \Attack{} and \Converge{} gate but disables adversarial hardening.
    \item \textbf{\sys.} The full protocol obtains a reproducing \pov $e_0$,
    requires the patcher to block it, and then
    attacks the patched program for distinct same-vulnerability bypasses. Each
    confirmed bypass becomes a permanent repair requirement, and the final patch
    must also pass differential regression checks. The main
    configuration uses $R{=}4$ hardening rounds.
    \item \textbf{Loop-depth variants.} These vary only the maximum number of
    adversarial rounds while keeping the same bootstrap, patching,
    execution-referee, and oracle machinery.
\end{itemize}
The arms therefore differ only in repair evidence and loop structure. Stage
prompts, model settings, execution contracts, and per-stage budgets are
identical wherever the corresponding stage is present.

\paperparagraph{Published baselines.}
We compare against published repair systems on shared subjects when patches or
reproduction artifacts are available. We reconstruct each patched tree from the
vulnerable revision and recorded diff, then score it under our held-out
fix-closed and residual suites. This does not claim prior systems failed under
their original criteria; it measures what their validation may not have tested.
\begin{itemize}[leftmargin=*]
    \item \textbf{\textsc{San2Patch}~\cite{kim2025san2patch}.}
    \textsc{San2Patch} is driven from the benchmark's own crashing input, never
    from our certified \povs, to avoid oracle leakage.
    \item \textbf{\textsc{PatchAgent}~\cite{yu2025patchagent}.}
    \textsc{PatchAgent} is evaluated under the same non-leakage rule, using the
    benchmark's available reproduction artifact rather than held-out \povs.
    \item \textbf{LoopRepair~\cite{ye2026looprepair}.} LoopRepair is re-run end
    to end on VulnLoc+ subjects using the same model as our runs, holding
    subject and agent capability fixed where possible.
\end{itemize}

\begin{table}[t]
\centering
\small
\caption{\bench summary: 101 CVEs, 61 repositories, and 20 weakness classes.
All subjects have fix-closed suites; 40 also have residual suites.}
\label{tab:corpus}

\begin{tabular*}{\columnwidth}{@{\extracolsep{\fill}}lrrr@{}}
\toprule
\headrow
                                & Java & C/C++ & All \\
\midrule
CVEs                            &  60 &  41 & 101 \\
CVEs with a fix-closed suite    &  60 &  41 & 101 \\
CVEs with a residual suite      &  28 &  12 &  40 \\
\midrule
Fix-closed \povs                & 213 & 111 & 324 \\
Residual \povs                  &  51 &  15 &  66 \\
\bottomrule
\end{tabular*}
\end{table}

\paperparagraph{Models and cost.}
The main comparison holds model capability fixed: all runs in
Sections~\ref{sec:rq1}--\ref{sec:rq3} use Claude Haiku 4.5. We additionally run
the full configuration on three unrelated model families to test model dependence
(Section~\ref{sec:rq4}). Reported cost is provider-billed where available and
otherwise estimated from the harness; we flag cases where fallback pricing
overstates third-party model costs.

\paperparagraph{Metrics and tests.}
Coverage is defined only for a patch. Paired comparisons therefore follow an
intention-to-treat policy: every subject carrying a certified suite is scored,
and a configuration that shipped no accepted patch there is credited zero.
For a certified suite $S$, coverage is the fraction of conclusive \povs blocked
by the generated patched program $P_{\mathrm{gen}}$:
\[
C(S,P_{\mathrm{gen}})=
\frac{
|\{e \in S : \textsc{Exec}(P_{\mathrm{gen}},e)=\textsc{Blocked}\}|
}{
|\{e \in S : \textsc{Exec}(P_{\mathrm{gen}},e)\in
\{\textsc{Blocked},\textsc{Repro}\}\}|
}.
\]
We report $\Cfix$ for fix-closed suites and $\Cres$
for residual suites; inconclusive executions are excluded from the denominator.
Certified suite
sizes vary from 1 to 14 \povs, so the primary metric is macro coverage, the
average of per-subject coverage; micro coverage pools all certified \povs as a
path-level diagnostic. Primary tests are exact Wilcoxon signed-rank
tests on per-subject deltas for $\Cfix$ and $\Cres$,
with Holm adjustment within primary and language-family groups~\cite{wilcoxon1945signedrank}. Language
splits, Section~\ref{sec:rq3} stratification, published-baseline
re-evaluation, and the model sweep are exploratory.

\section{Evaluation}
\label{sec:eval}

We evaluate whether adversarial hardening exposes bypasses that one-shot repair misses, whether those bypasses translate into higher held-out security coverage, and whether the effect generalizes across models and published repair systems.

We ask five questions. \textbf{RQ1} tests whether accepted repairs can still be
bypassed after blocking their construction \pov. \textbf{RQ2} measures
held-out coverage and cost relative to single-pass repair from the same finding.
\textbf{RQ3} attributes the gains to the bootstrap \pov and adversarial
hardening loop. \textbf{RQ4} tests model dependence. \textbf{RQ5} compares
published baselines under the same independent security oracles.

\subsection{Bypasses After Accepted Repair (RQ1)}
\label{sec:rq1}

Unless otherwise noted, all experiments use the setup of
Section~\ref{sec:setup}.

If blocking the \pov that triggered a patch were sufficient evidence of
repair, an attacker targeting that patched program would find no further
same-vulnerability path. We test this directly by turning the exploitation
agent against the patch produced from its own bootstrap \pov.

\textbf{Two runs in five expose a bypass after the patch has passed its own
gate.} In 41 of 101 runs (41\%), the exploiter produced a distinct,
execution-confirmed bypass of a patch that had already blocked the \pov it
was built from. Across those runs, the exploiter found 68 bypasses in total.
Ninety-nine runs reached the adversarial phase, so the rate over attackable
runs is also 41\%. The per-run distribution is 60 runs with no bypass, 26 with
one, 7 with two, 4 with three, and 4 with four. A system that stopped after the
first blocked \pov would have accepted all 41 bypassed patches without any
signal that the repair remained incomplete.

\textbf{The bypass rate does not decay with depth.} Conditioned on reaching
each hardening round, the probability of finding another confirmed bypass is
41\%, 37\%, 53\%, and 50\% over cohorts of 99, 41, 15, and 8 runs, respectively
(Figure~\ref{fig:attrition}). We do not interpret the later rates precisely
because the cohorts are small; the important point is that the rate does not
collapse after the first bypass. Four runs reached the four-round budget while
still confirming a new bypass in the final round, so search was truncated rather
than exhausted on those subjects.

The orchestrator enforces execution, not semantic novelty. A candidate
bypass counts only if execution reproduces the vulnerable behavior against the
current patch. The new bypass and $e_0$ are protected by the per-round gate;
earlier bypasses are replayed by the post-hardening gate described in
Section~\ref{sec:hardening}. The orchestrator rejects candidates that merely
replay a previous \pov command.
However, we prompt for semantic distinctness beyond that---whether a candidate
uses a genuinely different sink rather than a re-encoding of the same path, for
example---but do not machine-check it. The 41\% rate is therefore an upper bound on
semantically distinct bypasses. The key result is still execution-grounded:
patches that had passed their own construction-time checks often remained
vulnerable to additional same-vulnerability \povs.

\vspace{2pt}
\takeaway{Reusing construction evidence as the only validation signal is
insufficient: 41\% of runs produce a patch that blocks its bootstrap \pov
but is later bypassed by an execution-confirmed same-vulnerability \pov.}

\begin{table}[t]
\centering
\footnotesize
\caption{Primary held-out macro coverage under intention-to-treat scoring.
Paired columns report non-tied pairs ($k$), wins/ties/losses, and Holm-adjusted
exact Wilcoxon signed-rank $p$-values.}
\label{tab:coverage}

\setlength{\tabcolsep}{2pt}
\renewcommand{\arraystretch}{1.06}
\begin{tabular*}{\columnwidth}{@{\extracolsep{\fill}}lrrrrrrr@{}}

\toprule
\headrow
Oracle & $n$ & \multicolumn{2}{c}{Macro coverage} &
\multicolumn{4}{c}{Paired comparison} \\
\cmidrule(lr){3-4}\cmidrule(l){5-8}
\headrow
 & & \base & \sys & $k$ & $\Delta$ & $+/{=}/{-}$ &
 $p_{\mathrm{Holm}}$ \\
\midrule
$\Cfix$
  & 101 & 0.781 & \best{0.847} & 20
  & \gain{$+0.065$} & $17/81/3$ & \textbf{0.0286} \\
$\Cres$
  & 40 & 0.421 & \best{0.625} & 13
  & \gain{$+0.204$} & $11/27/2$ & \textbf{0.0254} \\
\bottomrule
\end{tabular*}
\end{table}

\subsection{Held-Out Coverage and Cost (RQ2)}
\label{sec:rq2}

RQ1 shows that accepted \pov-blocking patches can still be bypassed. RQ2
asks whether adversarial hardening improves final coverage on independent
oracles, and whether the improvement is large enough to justify its additional
cost.

\textbf{\sys produces accepted repairs on 98 of 101 subjects.} A patch is
accepted only after it builds, blocks every live \pov in the requirement set,
passes differential regression checks, and survives the standardized
manual-review protocol of Section~\ref{sec:manual-review}. Two failures occur
before repair: on CVE-2016-1833 the agent did not produce a working \pov, and
on libtiff bugzilla-2611 it produced a candidate \pov that did not trigger the
vulnerability within three attempts. The third failure is a patch rejected by
manual review. No run was lost at a post-patch execution gate: failed gates
returned to the responsible agent, and the correction loop converged within
the configured budget for every run that entered repair.

\textbf{Held-out coverage improves on both oracle families.}
Table~\ref{tab:coverage} reports the primary comparison: macro-averaged
coverage over every subject carrying a certified suite, with any subject that a
configuration left unrepaired credited zero. Fix-closed coverage rises from 0.781 to
0.847 over all 101 subjects, while beyond-upstream coverage rises from 0.421 to
0.625 over all 40 residual subjects. Both primary tests survive Holm
adjustment~\cite{holm1979stepdown}.

At the path level, micro coverage also increases: over the 324 certified
fix-closed paths, $\Cfix$ rises from 0.710 to 0.781, and over the
66 certified residual paths, $\Cres$ rises from 0.409 to 0.712.
Perfect coverage increases from 65 to 76 subjects for fix-closed suites and from
15 to 24 subjects for residual suites. The distribution is sparse but
directional. 81 of 101 subjects tie on $\Cfix$, so 20 discordant
pairs drive the fix-closed test. Among those pairs, the matched-pairs
rank-biserial correlation is 0.55 for $\Cfix$ and 0.76 for
$\Cres$~\cite{cureton1956rankbiserial,kerby2014rankbiserial}.
The number needed to treat---the number of subjects run through \sys per
net improved subject---is 7.2 for fix-closed coverage and 4.4 for residual
coverage~\cite{laupacis1988nnt}. Twenty-four of the 101 subjects improve on at
least one oracle and three regress on at least one. Thus, the gain is not
universal, but when the oracle distinguishes the two patches, the direction is
strongly favorable to \sys.

The full protocol is more expensive than single-pass repair, with a mean
per-subject cost of \$2.09 versus \$0.45 for \base, reflecting the additional
exploitation and correction stages.

\vspace{2pt}
\takeaway{\sys improves held-out coverage on both oracle families: $\Cfix$ rises by 0.065
over all 101 subjects and $\Cres$ by 0.204 over all 40 residual
subjects, and both primary tests survive Holm adjustment.}

\subsection{Attributing the Gains (RQ3)}
\label{sec:rq3}

\sys differs from \base in two ways: the patcher first sees a live
\pov, and the patched program is then attacked for bypasses. RQ3 separates
these effects using the PoV-only configuration, in which bootstrap \pov
generation and patch convergence are enabled but adversarial hardening is
disabled, and the loop-depth variants, which vary the number of hardening
rounds.

\textbf{The loop, not the \pov, carries the detectable gain.}
Table~\ref{tab:ablation} decomposes the total gain from
Table~\ref{tab:coverage}. On fix-closed paths, the live \pov contributes
$+0.027$ and the adversarial loop a further $+0.038$; on residual paths, the
\pov contributes $+0.050$ and the loop $+0.154$. The loop step is
statistically detectable on both oracles ($p=0.0020$ and $p=0.0156$), whereas
the \pov step is detectable on neither ($p=0.2944$ and $p=0.5000$).

The asymmetry reflects the intention-to-treat scoring policy. Requiring a live
\pov before patching makes the loop possible, but it also accounts for the
three subjects on which \sys ships no accepted patch: two runs never produce a
reproducing \pov, and a third ships a patch that manual review rejects. None of
the three reaches a hardening patcher, so the yield cost is charged to the
\pov step. The loop then improves 10 subjects on $\Cfix$ and 7 on $\Cres$
without lowering coverage on any subject. Residual paths separate the two
steps most strongly: fix-closed paths are aligned with the official fix and
often with the reported finding, whereas residual paths are precisely those
the official patch leaves open.

\textbf{Every coverage regression predates the loop.} The hardening loop is
$10/91/0$ on $\Cfix$ and $7/33/0$ on $\Cres$: it never
lowered a score. All three negative pairs in Table~\ref{tab:coverage} arise
before it. Two are the unrepaired subjects credited zero. The third is a genuine
coverage regression attributable to \pov anchoring: on CVE-2018-17297, the
bootstrap \pov led the patcher to repair one zip-slip path, while the
\base patch repaired a broader trace set. The exploiter found no bypass on
that subject, so the loop did not run and could not widen the repair.

Most residual benefit arrives early. The round-budget
sweep, reported in Table~\ref{tab:round-sweep}
and plotted in Figure~\ref{fig:attrition} (right), shows diminishing returns
after the first two hardening rounds. Moving from PoV-only to one
adversarial round yields the largest residual gain; later rounds add smaller
increments. This supports
using bounded hardening as a practical repair protocol rather than treating the
loop as an unbounded search for completeness.

\begin{table}[t]
\centering
\small
\caption{Attribution of the coverage gain to the bootstrap \pov and adversarial
loop. Each $\Delta$ is relative to the preceding row; $p$ is an exact Wilcoxon
signed-rank test.}
\label{tab:ablation}

\begin{tabular*}{\columnwidth}{@{\extracolsep{\fill}}lrrrr@{}}
\toprule
\headrow
Step & Mean & $\Delta$ & $+/{=}/{-}$ & $p$ \\
\midrule
\multicolumn{5}{l}{\emph{Fix-closed coverage} $\Cfix$, $n=101$} \\
\base                   & 0.781 & \na               & \na        & \na \\
+ live \pov             & 0.809 & \gain{$+0.027$}   & $10/88/3$ & 0.2944 \\
+ adversarial loop      & \best{0.847} & \gain{$+0.038$} & $10/91/0$ & \textbf{0.0020} \\
\addlinespace
\multicolumn{5}{l}{\emph{Beyond-upstream coverage} $\Cres$, $n=40$} \\
\base                   & 0.421 & \na               & \na        & \na \\
+ live \pov             & 0.471 & \gain{$+0.050$}   & $4/34/2$  & 0.5000 \\
+ adversarial loop      & \best{0.625} & \gain{$+0.154$} & $7/33/0$  & \textbf{0.0156} \\
\bottomrule
\end{tabular*}
\end{table}

\takeaway{Priced end to end, the bootstrap \pov pays for the loop rather
than for coverage: its own contribution is not statistically detectable on
either oracle once the runs it costs are charged to it. The loop is where the
detectable gain is --- $+0.038$ on $\Cfix$ and
$+0.154$ on $\Cres$, with no oracle regressions.}

\subsection{Model Dependence (RQ4)}
\label{sec:rq4}

An agentic repair protocol can easily become a report about one model. We
therefore run the full protocol with one model from each of four unrelated
families on the same 101 subjects, budgets, and held-out suites.

\textbf{Bypass discovery replicates across models.} The exploiter falsifies
an already-blocking patch in 47\%, 41\%, 35\%, and 31\% of runs across the four
models (Table~\ref{tab:models}). The RQ1 result is therefore not specific to any
one of the four tested model families. Because each arm uses the same model as
both exploiter and patcher, these rates characterize the full
repair trajectory rather than exploiter capability alone.

\textbf{The protocol transfers across models, while performance varies.} All
four models produce accepted repairs on a majority of subjects. Under
intention-to-treat scoring, the arms differ in both accepted-repair rate and
held-out coverage. Claude Haiku 4.5, the model held fixed in
Sections~\ref{sec:rq1}--\ref{sec:rq3}, produces an accepted repair on 98 of 101
subjects and reaches $\Cfix=0.847$ and $\Cres=0.625$; we use it as the reference
within the model sweep.

GLM 5.2 leads the sweep. It produces an accepted repair on 100 of 101 subjects
and reaches $\Cfix=0.921$ and $\Cres=0.795$, respectively 0.074 and 0.170 higher
than Haiku despite producing only two additional accepted repairs. The
difference is not solely an acceptance effect. On the 62 subjects for which
every model produces an accepted repair with a conclusive fix-closed score, GLM
remains highest at $\Cfix=0.953$; over the 22 of those subjects with a residual
suite, it also leads at $\Cres=0.786$. GLM produces the most execution-confirmed
bypasses---102 across 47\% of runs, compared with 68 across 41\% for Haiku.
Twelve GLM runs exhaust the four-round budget, compared with four Haiku runs,
and seven GLM subjects cost more than \$9; no Haiku subject does. Its \$2.43
mean cost is therefore the largest in the table even though its median subject
(\$1.66) is cheaper than Haiku's \$1.85. DeepSeek V4 Flash produces an accepted
repair on 94 subjects and reaches $\Cfix=0.821$ and $\Cres=0.608$, within three
points of Haiku on both oracles, at roughly a quarter of its cost.

\textbf{Strong accepted repairs do not compensate for unrepaired subjects.}
GPT-5.6-luna produces an accepted repair on only 66 subjects, giving it the
lowest intention-to-treat coverage in the sweep: $\Cfix=0.594$ and
$\Cres=0.400$.
Across those 66 accepted repairs, mean fix-closed coverage is $\Cfix=0.909$;
over the 24 accepted repairs on subjects with a residual suite, mean residual
coverage is $\Cres=0.667$. Both are second only to GLM. That accepted-only
result does not compensate for leaving a third of the benchmark unrepaired under
intention-to-treat scoring.
The shortfall is late rather than at the bootstrap: of the 35 unrepaired subjects, 26 have completed patches rejected under the
manual-review protocol of Section~\ref{sec:manual-review}, five exhaust an
execution gate's correction budget, and four never produce a working \pov.

\begin{table}[t]
\centering
\small
\caption{Full \sys protocol across four model families under intention-to-treat
scoring. ``acc.'' reports accepted repairs; ``byp.'' reports the share of runs
with an execution-confirmed bypass. Cost is the mean per subject and wall time
the median. $\Cfix$ and $\Cres$ use all 101 and all 40 subjects for every arm.}
\label{tab:models}
\setlength{\tabcolsep}{4pt}
\begin{tabular}{@{}lrrrrrr@{}}
\toprule
\headrow
Model & acc. & byp. & $\Cfix$ & $\Cres$ & \$/subj. & min. \\
\midrule
Claude Haiku 4.5  & 98 & 41\% & 0.847 & 0.625 & 2.09 & 25.8 \\
GLM 5.2           & 100 & 47\% & \best{0.921} & \best{0.795} & 2.43 & 37.1 \\
DeepSeek V4 Flash & 94 & 35\% & 0.821 & 0.608 & 0.54 & 37.5 \\
GPT-5.6-luna      & 66 & 31\% & 0.594 & 0.400 & 0.16 & 11.8 \\
\bottomrule
\end{tabular}
\vspace{2pt}

\end{table}

\takeaway{Bypass discovery replicates across all four tested model families,
and every arm produces accepted repairs on a majority of subjects. End-to-end
performance nonetheless differs: GLM 5.2 produces the most accepted repairs
and the highest held-out coverage, but also has the highest mean cost and 12
runs that exhaust the four-round budget. GPT-5.6-luna produces high coverage
among its accepted repairs, but leaves 35 of 101 subjects unrepaired.}

\subsection{Published Baselines (RQ5)}
\label{sec:rq5}

The certified suites do not depend on \sys, so they can also score patches
produced by other repair systems. We compare \sys with \base on the full
benchmark, and with \textsc{San2Patch},
LoopRepair, and \textsc{PatchAgent} on pairwise shared subjects. Within each
comparison, both systems are scored on the same CVEs, under the same oracle
suites and the same intention-to-treat policy used everywhere else in this
section. A system that reported no patch, or whose own pipeline rejected one,
is credited zero on that subject rather than dropped. These comparisons are still not a global
ranking: the subject set, input representation, and available artifacts
differ by system.

The \base comparator is the single-pass arm from Sections~\ref{sec:rq2}--\ref{sec:rq4}:
the same patching agent receives the normalized finding but no
live \pov and no post-patch bypass search. All five systems evaluated in this section---\base, \sys, \textsc{San2Patch}, LoopRepair, and \textsc{PatchAgent}---use Claude Haiku 4.5.

\textbf{\sys has higher held-out coverage in every pairwise comparison.}
Table~\ref{tab:baselines} shows higher $\Cfix$ and $\Cres$ on every shared
subject set. The largest fix-closed separation is against LoopRepair
($+0.605$), while residual gains range from $+0.454$ to $+0.500$ across the
published systems.

Independent oracles expose gaps that original validation did not test.
LoopRepair reports a patch on 30 of 40 subjects, each accepted by its own
single-crashing-input verification. Under our certified suites only 10 of 39
measurable shared subjects block every fix-closed path. \textsc{San2Patch}
and \textsc{PatchAgent} also improve some subjects but leave residual and
fix-closed paths open under independent oracles. This does not contradict
their original evaluations; it shows their validation criteria did not
measure the same held-out attack-path coverage.

Table~\ref {tab:baseline-cost} reports mean operational cost per run.
\sys is consistently more expensive because it performs
additional exploitation and correction stages after the initial repair. Accordingly, these comparisons should be interpreted as an end-to-end
coverage--cost tradeoff rather than an efficiency ranking: the systems differ
in their inputs, repair protocols, and validation policies.

\begin{table}[t]
\centering
\small
\caption{Pairwise held-out coverage on shared subjects under identical certified
suites and intention-to-treat scoring. $n$ is the number of shared subjects with
a measurable suite; ``Comp.'' denotes the comparison system.}
\label{tab:baselines}
\begin{tabular*}{\columnwidth}{@{\extracolsep{\fill}}llrrrr@{}}
\toprule
\headrow
System & Oracle & $n$ & Comp. & \sys & $\Delta$ \\
\midrule
\base               & $\Cfix$ & 101 & 0.781 & \best{0.847} & \gain{$+0.065$} \\
                    & $\Cres$ &  40 & 0.421 & \best{0.625} & \gain{$+0.204$} \\
\addlinespace
\textsc{San2Patch}  & $\Cfix$ & 33 & 0.608 & \best{0.922} & \gain{$+0.314$} \\
                    & $\Cres$ & 10 & 0.100 & \best{0.600} & \gain{$+0.500$} \\
\addlinespace
LoopRepair          & $\Cfix$ & 39 & 0.325 & \best{0.930} & \gain{$+0.605$} \\
                    & $\Cres$ & 11 & 0.182 & \best{0.636} & \gain{$+0.454$} \\
\addlinespace
\textsc{PatchAgent} & $\Cfix$ & 16 & 0.836 & \best{0.929} & \gain{$+0.093$} \\
                    & $\Cres$ &  6 & 0.167 & \best{0.667} & \gain{$+0.500$} \\
\bottomrule
\end{tabular*}
\end{table}

\takeaway{On each shared subject set, \sys scores higher under the same
independent oracle suites than the evaluated comparator. The comparison is
end-to-end: it reflects differences in inputs, protocol, and validation, not
only the adversarial loop.}

\section{Discussion and Limitations}
\label{sec:discussion}

\paperparagraph{Bounded hardening, not proof.}
\sys does not prove that a generated patch is complete. The exploiter is a
budgeted agent, the live \pov set is finite, and the held-out suites contain
human-authored attack paths rather than the full input space. A stable run
therefore means only that the configured exploiter failed to find another
execution-confirmed same-vulnerability bypass and that the final patch covered
the certified suites to the reported degree~\cite{klees2018evaluating}. This is stronger than reusing the
construction \pov as the sole validation oracle, but it remains a testing
claim rather than a formal proof~\cite{le2019reliability}.

\paperparagraph{Oracle curation.}
Oracle certification is execution-based: a \pov enters a suite only by
reproducing on the recorded vulnerable tree and then satisfying the
fix-closed or residual certification rule. Because all configurations are
scored by identical held-out suites, oracle curation does not differentially
favor one arm within a paired comparison. The remaining risk is coverage bias:
the suites contain human-authored paths, not all relevant paths. Residual \povs
require particular care. They demonstrate same-vulnerability paths that survive
the official fix, but the official fix itself does not certify them as
fixed. Later upstream fixes corroborate 14 residual gaps
(Appendix~\ref{app:upstream}); for 8 of them we confirmed by execution that
the \pov is blocked on the later tree, and extending that to every row
remains future work.

\paperparagraph{Published-system comparison.}
The published-system comparison is an end-to-end oracle comparison, not a claim
that prior systems failed under their original criteria. Section~\ref{sec:rq5}
scores available patches on shared subjects using the same certified suites,
which measures held-out attack-path coverage that the original validations may
not have tested. The comparison therefore reflects differences in input
modality, repair protocol, repair budget, implementation maturity, and
validation policy, not only the adversarial loop. Its purpose is to show that
independent security oracles expose gaps missed by conventional validation, not
to produce a universal ranking of repair systems.

\paperparagraph{Model memorization.}
Pretraining exposure to public CVEs, advisories, fixes, or exploit examples
cannot be excluded~\cite{ramos2025memorizing}. Such exposure may inflate absolute coverage for every
configuration. It is less able to explain paired differences between
configurations that use the same model and differ only in the repair-time
evidence constructed during the run. It remains, however, a limitation of
evaluating LLM-based repair on public vulnerabilities. Redacting CVE identifiers, upstream
patches, advisory exploit text, and held-out \povs reduces direct leakage
through the prompt, but cannot remove knowledge already encoded in the model~\cite{liang2026swebenchillusion}.

\vspace{1mm}

\section{Related Work}

\paperparagraph{Automated vulnerability repair.}
Automated vulnerability repair spans learning-based, analysis-based, and
LLM-based systems~\cite{hou2024llmse}. Recent systematizations organize the area around
localization, patch generation, and validation, and identify correctness
assessment as a persistent bottleneck~\cite{hu2025sok,li2025effectivesok}.
Learning-based approaches often translate vulnerable code into fixed code under
an assumed edit location~\cite{ye2026looprepair,chen2021sequencer,jiang2021cure,wu2023vjbench}. Analysis-based systems such
as ExtractFix and VulnFix derive repair constraints from concrete failures or
vulnerability witnesses~\cite{gao2021extractfix,zhang2022vulnfix,mechtaev2016angelix,nguyen2013semfix}. Recent
agentic systems, including \textsc{Appatch}, \textsc{PatchAgent},
\textsc{San2Patch}, \textsc{CodeCureAgent}, the OSS-Fuzz repair agent, and
LoopRepair, use LLMs with vulnerable code, warnings, traces, or reproduction
artifacts and iterate on compiler, test, sanitizer, or reviewer
feedback~\cite{nong2025appatch,yu2025patchagent,kim2025san2patch,joos2026codecureagent,zhang2026ossfuzz,ye2026looprepair,zhang2026aixcc}.

\begin{table}[t]
\centering
\footnotesize
\caption{Operational repair cost over every actual run, including failed-run spend. All five methods were executed in our evaluation with Claude Haiku 4.5. PatchAgent cost is reconstructed from execution traces and is therefore a lower bound.}
\label{tab:baseline-cost}

\begin{tabular*}{\columnwidth}{@{\extracolsep{\fill}}lrrr@{}}
\toprule
\headrow
Method & Runs & Total (\$) & Mean/run (\$) \\
\midrule
\base               & 101 & 45.90  & 0.45 \\
\textsc{San2Patch}  & 38  & 18.22  & 0.48 \\
LoopRepair          & 40  & 15.03  & 0.38 \\
\textsc{PatchAgent} & 15  & 4.79   & 0.32 \\
\sys                & 101 & 210.76 & 2.09 \\
\bottomrule
\end{tabular*}
\end{table}

\paperparagraph{What repair loops optimize.}
Most recent systems iterate, but the iteration usually searches for a patch
against a fixed target: a warning, crash, sanitizer trace, supplied \pov, or
reviewer failure~\cite{zhang2024autocoderover}. \textsc{San2Patch} explores reasoning paths,
\textsc{PatchAgent} iterates over tool actions and rejected patches,
LoopRepair ranks candidate patches using taint-trace feedback, and
\textsc{CodeCureAgent} iterates through tool calls and approver
feedback~\cite{kim2025san2patch,yu2025patchagent,ye2026looprepair,joos2026codecureagent}.
\sys changes a different object: the repair objective. Once a patch blocks the
current \povs, the patched program becomes the next attack target, and each
execution-confirmed same-vulnerability bypass becomes a new repair requirement.
This resembles counterexample-guided synthesis~\cite{solar2006combinatorial,clarke2000cegar},
but uses a budgeted exploiter rather than a complete verifier.

\paperparagraph{Exploit and \pov generation.}
A complementary line of work generates proof-of-vulnerability artifacts rather
than patches~\cite{avgerinos2011aeg,cha2012mayhem}. SemFuzz generates Linux-kernel exploits from vulnerability text
and code semantics~\cite{you2017semfuzz}; FaultLine generates language-agnostic
\pov tests~\cite{nitin2025faultline}; PoC-Gym studies static-trace guidance for
Java \povs~\cite{gezgin2026pocgym}; and CVE-Genie reconstructs environments
and verifiable exploits from public CVE material~\cite{ullah2025cvegenie}.
These systems focus on making a vulnerability executable~\cite{wang2018revery}. \sys uses that
capability as a repair primitive: the first \pov bootstraps repair, while
later \povs test whether the current patch can be bypassed.

\paperparagraph{Patch validation and overfitting.}
Patch validation is a long-standing weakness of automated repair~\cite{legoues2019apr,monperrus2018bibliography,wang2020apca}. Test-suite
overfitting can produce plausible patches that pass available tests but miss the
intended repair~\cite{smith2015overfitting,qi2015plausibility}, and generated tests can expose
such patches~\cite{yang2017bettertests,xin2017difftgen}. Security repair inherits this problem:
blocking one \pov, sanitizer trace, or functional test does not imply that the
security boundary is enforced. PVBench shows that \pov-plus-functional-test
validation can overestimate vulnerability-repair success~\cite{yu2026pvbench},
while VulnRepairEval and PatchEval move toward executable, containerized
security tests for LLM patch assessment~\cite{wang2025vulnrepaireval,wei2025patcheval}.
\sys adopts executable validation but separates construction-time \povs
from held-out evaluation suites.

\paperparagraph{Benchmarks, official fixes, and residual risk.}
Many repair evaluations use the developer or official fix as the target
behavior, through syntactic matching, supplied tests, or paths known to be
closed by the recorded fix~\cite{gao2021extractfix,zhang2022vulnfix,yu2026pvbench,wei2025patcheval,bhuiyan2023secbenchjs,bui2022vul4j,mei2026arvo}.
This is useful, but it can make the official fix a ceiling: paths the official
fix leaves open are not measured. Prior work shows that security patches can be
incomplete or require later correction~\cite{li2017securitypatches,gu2010bugfixed,liu2025incompletefixes,park2012supplementary}, and that
vulnerable code propagates across projects through reuse~\cite{reid2022orphan}. \sys
therefore reports fix-closed \povs, which are blocked by the official fix,
separately from residual \povs, which remain reproducible after the official
fix and measure beyond-upstream hardening.

\vspace{1mm}

\section{Conclusion}

This paper presented \sys, an exploit-guided vulnerability-repair protocol that
treats each accepted patch as a new attack target. Rather than validating a
repair only against the evidence that produced it, \sys grows the repair
objective whenever execution confirms a new same-vulnerability bypass and
evaluates the final patch with independent held-out security oracles. Across
101 C/C++ and Java CVE subjects, the exploiter bypassed already-accepted
patches in 41\% of runs. Relative to single-pass repair from the same finding,
\sys increased held-out fix-closed coverage from 0.781 to 0.847 and residual
coverage from 0.421 to 0.625. Re-scoring published repair systems under the
same certified oracles further showed that patches accepted by their original
validation can leave attack paths unblocked. These results suggest that
adaptive counterexample growth and independent evaluation provide a stronger
basis for assessing automated vulnerability repair, while remaining a bounded
testing methodology rather than a proof of patch completeness.

\newpage
\bibliographystyle{plain}
\bibliography{references-v5,references-extra}
\clearpage

\appendix
\raggedbottom

\section{Ethical Considerations}
\label{sec:ethics}

This work studies automated repair for already reported vulnerabilities, but it
has dual-use risk. The same mechanism that tests whether a patch is incomplete
can generate executable proof-of-vulnerability artifacts. We therefore designed
the evaluation to reduce harm while preserving scientific auditability. All
experiments run on pinned historical revisions inside project containers or
per-run worktrees. Agents do not scan live services, interact with third-party
deployments, or retrieve web content during repair. Their inputs exclude CVE
identifiers, upstream patches, advisory exploit text, and held-out evaluation
\povs, and all generated commands, edits, and model outputs are treated as
untrusted artifacts.

The most sensitive cases are residual \povs: \povs that reproduce both
before and after the recorded official fix. Such artifacts can indicate that a
released fix left a same-vulnerability path open, and in some cases our
upstream-history analysis found that maintainers later closed the same behavior.
We therefore do not treat residual \povs as ordinary benchmark tests. For
residual paths that remain open in maintained projects, we follow coordinated
disclosure and withhold the executable artifacts (payload, harness, and
command) until the affected project ships a fix; public materials describe only
the weakness class, certification rule, and aggregate outcome.

\legacyparagraph{Responsible disclosure and reporting.}
We have reported all 9 gaps that remain open at upstream's current head
(Appendix~\ref{app:upstream}) to the 8 affected projects, each through its
stated security channel: the ASF security team for the Apache subjects, a
private GitHub advisory or a vendor security address otherwise. Each report
carries the minimum needed to triage: affected component and revision,
weakness class, certification rule, and a summary of the surviving path; full
reproduction details are available to each project on request. Two of the repositories have been dormant for more than two
years, so those notifications are best-effort.

\vspace{1mm}

Historical residual paths that are already fixed in current upstream releases
are released with the artifact. They serve as evidence that the residual oracle
can identify security-relevant gaps.

Our claims are bounded accordingly. \sys is a repair-evaluation method, not a
tool for attacking deployed systems. A residual \pov shows that a
particular behavior remains reproducible under our benchmark contract; it does
not by itself prove exploitability in every deployment or replace maintainer
review. Conversely, withholding sensitive \povs limits immediate
reproducibility for those cases, but avoids increasing risk to users of projects
whose fixes may still be incomplete.

\section{Open Science and Artifact Release}
\label{sec:openscience}

\legacyparagraph{Artifact availability.} The artifact is publicly available at
\url{https://github.com/P2Patch/P2patch}. It contains the orchestration code,
configuration files, agent prompt templates, and scoring scripts, together with
\bench: each subject's pinned revision, container definition, and certified
\pov suites, as well as the harness used to re-score the published baselines.
The pipeline runs behind the reported numbers are distributed separately and
linked from the repository.

\legacyparagraph{Withheld artifacts.} Following the disclosure policy of
Appendix~\ref{sec:ethics}, we withhold the executable \povs for residual gaps
that are not confirmed closed upstream: 18 of the 40 residual suites (30 of 66
\povs), together with 4 fix-closed suites whose harnesses also exercise a
withheld path (16 of 324 \povs). All other suites are released. A withheld
suite is published once the affected project ships a fix, or once its
maintainers agree that release poses no risk to users; until then, maintainers
and researchers can request access. Per-subject scores for every subject,
including the withheld ones, are reported in Tables~\ref{tab:persubject}
and~\ref{tab:persubjectc}.

\section{Protocol and Orchestration Details}
\label{app:protocol}

This appendix gives the stage-level protocol used by the orchestrator. The main
paper describes the design invariants; Algorithm~\ref{alg:protocol} records the
exact control flow used in the evaluation, including bootstrap attempts,
patch-convergence attempts, adversarial hardening rounds, and post-hardening replay.

A run returns a patch only if all construction-time gates succeed. Any bootstrap
failure, exhausted correction budget, failed post-hardening replay, or inconclusive
protected \pov produces a rejected repair, which receives no coverage credit
under the intention-to-treat policy. The evaluation uses $\katt{=}3$ attacker attempts, $\kgate{=}3$ gate evaluations per convergence call, and $R{=}4$ adversarial rounds unless
otherwise stated.

\begin{algorithm}[!t]
\footnotesize
\DontPrintSemicolon
\caption{Bounded exploit--patch protocol used in the evaluation. The
orchestrator decides every transition through execution; neither agent certifies
its own success. Any $\bot$, crashed agent, or refusal stops the run and is
recorded as a rejected repair.}
\label{alg:protocol}
\KwIn{pinned vulnerable revision $P_0$; normalized finding $f$; budgets $\katt$
(attacker attempts), $\kgate$ (gate evaluations), $R$ (hardening rounds)}
\KwOut{hardened program $P$ with requirement set $E$, or $\bot$}
\KwData{$\Exec(P,e)\in\{\repro,\blocked,\incon\}$ is the referee of
Section~\ref{sec:referee}; $\blk{e}$ abbreviates $\Exec(P,e)=\blocked$ on the
current $P$; $\noreg$ holds when no command in the append-only set $T$ regresses
against the pristine baseline.}
\BlankLine
isolate a worktree at $P_0$; build the container\;
\tcp{Bootstrap: make the finding executable}
\For{$a \leftarrow 1$ \KwTo $\katt$}{
  $e_0 \leftarrow$ \Attack{$f$, $P_0$, $\emptyset$}\;
  \lIf{$\Exec(P_0, e_0) = \repro$}{\KwSty{break}}
  \lIf{$a = \katt$}{\Return{$\bot$}}
}
$E \leftarrow \{e_0\}$\;
\BlankLine
\tcp{First patch: the patcher sees the \pov}
$P \leftarrow$ \Patch{$P_0$, $f$, $e_0$}\;
$P \leftarrow$ \Converge{$P$, $[\blk{e_0}]$}\;
\BlankLine
\tcp{Hardening: the attacker sees the diff}
\For{$r \leftarrow 1$ \KwTo $R$}{
  $e_r \leftarrow$ \Attack{$f$, $P$, $e_0$}\;
  \lIf{no candidate is proposed}{\KwSty{break}}
  \lIf{$\Exec(P, e_r) \neq \repro$}{\KwSty{break}}
  $E \leftarrow E \cup \{e_r\}$\tcp*{monotone}
  $P \leftarrow$ \Patch{$P$, $e_r$}\;
  $P \leftarrow$ \Converge{$P$, $[\blk{e_r}, \blk{e_0}]$}\;
}
\BlankLine
\tcp{Terminal replay: no correction attempt}
\ForEach{$e \in E \setminus \{e_0\}$}{
  \lIf{$\Exec(P, e) \neq \blocked$}{\Return{$\bot$}}
}
\tcp{Regression gate: may revise $P$}
$P \leftarrow$ \Converge{$P$, $[\blk{e_0}, \noreg]$}\;
\Return{$(P, E)$}\tcp*{held-out suites score $P$}
\BlankLine
\Fn{\Converge{$P$, predicate list $\Phi$}}{
  \For{$c \leftarrow 1$ \KwTo $\kgate$}{
    $\varphi \leftarrow$ first predicate in $\Phi$ failing on $P$\;
    \lIf{there is no such $\varphi$}{\Return{$P$}}
    \lIf{$c = \kgate$}{\Return{$\bot$}}
    $P \leftarrow$ \Patch{$P$, $\varphi$, evidence for $\varphi$}\;
  }
}
\end{algorithm}

\section{Benchmark and Oracle Construction}
\label{app:benchmark}

This appendix section describes how subjects, findings, execution contracts,
and held-out oracle suites are assembled. The main text reports the compact
benchmark summary; here we make the construction process auditable.

\legacyparagraph{Subject construction.}
Each subject consists of a pinned vulnerable revision, a normalized finding, a
container definition, project build commands, PoV-execution commands, and
regression commands when available. C/C++ subjects are derived from VulnLoc+
subjects whose required contracts could be reproduced. Java subjects are
curated from GitHub Security Advisories and NVD entries and paired with
manually checked CodeQL-derived taint findings.

\legacyparagraph{Oracle construction.}
Held-out PoVs are authored outside the repair loop and finalized before scoring.
Agents never receive the oracle source files, commands, expected outcomes,
certification results, CVE identifier, advisory exploit text, or upstream patch.
A fix-closed PoV is retained only if it reproduces on the vulnerable revision
and is blocked by the official fix. A residual PoV is retained only if it
reproduces on both the vulnerable revision and the official fixed revision.

\legacyparagraph{Benchmark flow.}
\bench is assembled from 138 authored findings, of which 101 reached a
reproducible subject and a certified fix-closed suite. Those 101 are the scored
subject set of Table~\ref{tab:corpus}; the remaining 37 are not scored.
Residual suites exist only where the official fix leaves a demonstrated path
open, so 40 of the 101 carry one.

\vspace{1mm}

\begin{table*}[!t]
\centering
\footnotesize

\caption{All 101 \bench subjects (60 Java and 41 C/C++) with CWE and certified
\pov-suite sizes. $|\Sfix|$ and $|\Sres|$ are the fix-closed and residual suite
sizes; \na{} denotes no certified residual suite.}

\label{tab:benchmark}

\scriptsize
\renewcommand{\arraystretch}{1.05}

\begin{tabular*}{0.475\textwidth}{@{\extracolsep{\fill}}lllrr@{}}
\toprule
\headrow
CVE & Project & CWE & \multicolumn{2}{c}{Certified \povs} \\
\cmidrule(l){4-5}
 & & & $|\Sfix|$ & $|\Sres|$ \\
\midrule
2014-7816 & undertow & CWE-22 & 3 & \na \\
\rowcolor{p2gray}2016-5394 & sling-xss & CWE-79 & 3 & 2 \\
2016-6812 & cxf & CWE-79 & 2 & 1 \\
\rowcolor{p2gray}2016-9177 & spark & CWE-22 & 2 & 2 \\
2017-1000487 & plexus-utils & CWE-78 & 8 & \na \\
\rowcolor{p2gray}2018-1000129 & jolokia & CWE-79 & 6 & \na \\
2018-1002200 & plexus-archiver & CWE-22 & 4 & 2 \\
\rowcolor{p2gray}2018-1002201 & zt-zip & CWE-22 & 4 & \na \\
2018-1002202 & zip4j & CWE-22 & 3 & 2 \\
\rowcolor{p2gray}2018-1047 & wildfly & CWE-22 & 3 & 2 \\
2018-12036 & DependencyCheck & CWE-22 & 6 & 2 \\
\rowcolor{p2gray}2018-12542 & vertx-web & CWE-22 & 3 & \na \\
2018-1260 & spring-security-oauth & CWE-94 & 2 & \na \\
\rowcolor{p2gray}2018-17297 & hutool & CWE-22 & 5 & 2 \\
2018-9159 & spark & CWE-22 & 5 & \na \\
\rowcolor{p2gray}2019-0207 & tapestry-5 & CWE-22 & 2 & \na \\
2019-0225 & jspwiki & CWE-22 & 1 & 1 \\
\rowcolor{p2gray}2019-10077 & jspwiki & CWE-79 & 3 & \na \\
2019-17573 & cxf & CWE-79 & 6 & 1 \\
\rowcolor{p2gray}2020-2261 & perfecto-plugin & CWE-78 & 1 & 1 \\
2020-35460 & mpxj & CWE-22 & 4 & \na \\
\rowcolor{p2gray}2020-8570 & kubernetes-java & CWE-22 & 2 & \na \\
2021-29425 & commons-io & CWE-22 & 4 & \na \\
\rowcolor{p2gray}2021-41269 & cron-utils & CWE-94 & 2 & \na \\
2022-22932 & karaf & CWE-22 & 3 & \na \\
\rowcolor{p2gray}2022-22947 & spring-cloud-gateway & CWE-94 & 3 & \na \\
2022-23082 & curekit & CWE-22 & 2 & \na \\
\rowcolor{p2gray}2022-25842 & one-java-agent & CWE-22 & 3 & \na \\
2022-26049 & goomph & CWE-22 & 2 & \na \\
\rowcolor{p2gray}2022-26884 & dolphinscheduler & CWE-22 & 7 & 1 \\
2022-28367 & antisamy & CWE-79 & 2 & 2 \\
\rowcolor{p2gray}2022-32287 & uima-uimaj & CWE-22 & 3 & \na \\
2022-4244 & plexus-utils & CWE-22 & 3 & 1 \\
\rowcolor{p2gray}2022-42889 & commons-text & CWE-94 & 12 & \na \\
2022-44262 & ff4j & CWE-94 & 10 & \na \\
\rowcolor{p2gray}2022-46166 & spring-boot-admin & CWE-94 & 14 & \na \\
2022-46907 & jspwiki & CWE-79 & 7 & 1 \\
\rowcolor{p2gray}2022-4963 & spring-module-core & CWE-89 & 3 & \na \\
2023-24422 & script-security-plugin & CWE-78 & 3 & \na \\
\rowcolor{p2gray}2023-37460 & plexus-archiver & CWE-22 & 2 & \na \\
2023-37908 & xwiki-rendering & CWE-79 & 3 & 2 \\
\rowcolor{p2gray}2023-45277 & yamcs & CWE-22 & 6 & 5 \\
2023-45278 & yamcs & CWE-22 & 5 & 5 \\
\rowcolor{p2gray}2025-0851 & djl & CWE-22 & 2 & 2 \\
2025-1391 & keycloak & CWE-284 & 3 & \na \\
\rowcolor{p2gray}2025-1584 & solon & CWE-23 & 1 & 2 \\
2025-23026 & jte & CWE-79 & 4 & 2 \\
\rowcolor{p2gray}2025-26795 & iotdb & CWE-200 & 3 & \na \\
2025-26864 & iotdb & CWE-200 & 1 & \na \\
\rowcolor{p2gray}2025-27496 & snowflake-jdbc & CWE-532 & 1 & 1 \\
2025-27508 & emissary & CWE-327 & 4 & 2 \\
\bottomrule
\end{tabular*}
\hspace{5mm}
\begin{tabular*}{0.475\textwidth}{@{\extracolsep{\fill}}lllrr@{}}
\toprule
\headrow
CVE & Project & CWE & \multicolumn{2}{c}{Certified \povs} \\
\cmidrule(l){4-5}
 & & & $|\Sfix|$ & $|\Sres|$ \\
\midrule
2025-30067 & kylin & CWE-94 & 2 & 3 \\
\rowcolor{p2gray}2025-32897 & incubator-seata & CWE-502 & 2 & 1 \\
2025-32960 & restapi & CWE-79 & 2 & 1 \\
\rowcolor{p2gray}2025-32961 & jpawebapi & CWE-79 & 1 & 1 \\
2025-46096 & solon & CWE-22 & 3 & \na \\
\rowcolor{p2gray}2025-4641 & webdrivermanager & CWE-611 & 1 & \na \\
2025-47771 & powsybl-core & CWE-502 & 1 & \na \\
\rowcolor{p2gray}2025-52888 & allure2 & CWE-611 & 3 & \na \\
2025-53689 & jackrabbit & CWE-611 & 2 & 1 \\
\rowcolor{p2gray}2012-2806 & libjpeg-turbo & CWE-787 & 1 & \na \\
2012-5134 & libxml2 & CWE-119 & 1 & \na \\
\rowcolor{p2gray}2013-7437 & potrace & CWE-190 & 6 & 1 \\
2016-10092 & libtiff & CWE-119 & 1 & \na \\
\rowcolor{p2gray}2016-10272 & libtiff & CWE-119 & 1 & \na \\
2016-1833 & libxml2 & CWE-125 & 4 & \na \\
\rowcolor{p2gray}2016-1838 & libxml2 & CWE-125 & 1 & \na \\
2016-1839 & libxml2 & CWE-125 & 4 & \na \\
\rowcolor{p2gray}2016-3186 & libtiff & CWE-119 & 2 & \na \\
2016-5314 & libtiff & CWE-787 & 2 & 1 \\
\rowcolor{p2gray}2016-5321 & libtiff & CWE-119 & 2 & 1 \\
2016-5844 & libarchive & CWE-190 & 2 & \na \\
\rowcolor{p2gray}2016-8691 & jasper & CWE-369 & 2 & \na \\
2016-9264 & libming & CWE-119 & 3 & 1 \\
\rowcolor{p2gray}2016-9532 & libtiff & CWE-125 & 1 & \na \\
2016-9557 & jasper & CWE-190 & 1 & \na \\
\rowcolor{p2gray}2017-14745 & binutils-gdb & CWE-190 & 2 & \na \\
2017-15020 & binutils-gdb & CWE-125 & 3 & \na \\
\rowcolor{p2gray}2017-15025 & binutils-gdb & CWE-369 & 1 & 1 \\
2017-15232 & libjpeg-turbo & CWE-476 & 7 & \na \\
\rowcolor{p2gray}2017-5225 & libtiff & CWE-119 & 2 & 1 \\
2017-5969 & libxml2 & CWE-476 & 1 & 1 \\
\rowcolor{p2gray}2017-5974 & zziplib & CWE-119 & 1 & \na \\
2017-5975 & zziplib & CWE-787 & 2 & 1 \\
\rowcolor{p2gray}2017-5976 & zziplib & CWE-787 & 2 & 1 \\
2017-6965 & binutils-gdb & CWE-119 & 4 & \na \\
\rowcolor{p2gray}2017-7595 & libtiff & CWE-369 & 3 & \na \\
2017-7599 & libtiff & CWE-20 & 8 & \na \\
\rowcolor{p2gray}2017-7600 & libtiff & CWE-20 & 8 & \na \\
2017-7601 & libtiff & CWE-20 & 5 & 1 \\
\rowcolor{p2gray}2018-18557 & libtiff & CWE-787 & 1 & \na \\
2018-8806 & libming & CWE-416 & 6 & \na \\
\rowcolor{p2gray}2018-8964 & libming & CWE-416 & 5 & 4 \\
2020-27828 & jasper & CWE-787 & 2 & \na \\
\rowcolor{p2gray}2021-3272 & jasper & CWE-125 & 2 & \na \\
2022-4645 & libtiff & CWE-125 & 3 & \na \\
\rowcolor{p2gray}2022-48281 & libtiff & CWE-787 & 1 & \na \\
bz2611 & libtiff & CWE-369 & 2 & 1 \\
\rowcolor{p2gray}bz2633 & libtiff & CWE-125 & 1 & \na \\
gnu19784 & coreutils & CWE-125 & 3 & \na \\
\rowcolor{p2gray}gnu25023 & coreutils & CWE-125 & 2 & \na \\
\bottomrule
\end{tabular*}

\end{table*}

\begin{table}[t]
\centering
\footnotesize
\caption{Exploratory language breakdown of Table~\ref{tab:coverage} under the
same intention-to-treat scoring. Coverage is macro-averaged per subject; these
rows are descriptive diagnostics, not separate inferential results.}

\label{tab:coverage-lang}

\begin{tabular*}{\columnwidth}{@{\extracolsep{\fill}}llrrrr@{}}
\toprule
\headrow
Oracle & Subset & $n$ & \base & \sys & $\Delta$ \\
\midrule
$\Cfix$ & Java  & 60 & 0.730 & 0.801 & \gain{$+0.071$} \\
        & C/C++ & 41 & 0.857 & 0.913 & \gain{$+0.057$} \\

\addlinespace
$\Cres$ & Java  & 28 & 0.494 & 0.643 & \gain{$+0.149$} \\
        & C/C++ & 12 & 0.250 & 0.583 & \gain{$+0.333$} \\

\bottomrule
\end{tabular*}
\end{table}

\begin{table}[b]
\centering
\footnotesize
\caption{Round-budget sweep. Each row reports held-out coverage at the given
maximum number of adversarial rounds. The analysis includes subjects measured
at every budget: 98 for $\Cfix$ and 39 for $\Cres$. Within this set, the
intention-to-treat policy applies, so a run that shipped nothing is scored as
zero. The different denominators are why $R=4$ here sits marginally under
Table~\ref{tab:coverage}'s 0.847 and 0.625.}
\label{tab:round-sweep}

\begin{tabular*}{\columnwidth}{@{\extracolsep{\fill}}lrrrr@{}}
\toprule
\headrow
Budget & $\Cfix$ & $\Delta$ & $\Cres$ & $\Delta$ \\
\midrule
$R=0$ (\pov only) & 0.811 & \na      & 0.457 & \na \\
$R=1$             & 0.833 & \gain{$+0.022$} & 0.556 & \gain{$+0.099$} \\
$R=2$             & 0.848 & \gain{$+0.014$} & 0.590 & \gain{$+0.033$} \\
$R=3$             & 0.850 & \gain{$+0.002$} & 0.615 & \gain{$+0.026$} \\
$R=4$ (default)   & 0.850 & \gain{$+0.000$} & 0.615 & \gain{$+0.000$} \\
\bottomrule
\end{tabular*}
\end{table}

\section{Comparison with Prior Repair Systems}
\label{app:prior-systems}

This appendix collects the empirical and conceptual comparisons with prior
repair systems. Section~\ref{app:published-details} gives the empirical
re-scoring details: Table~\ref{tab:percase} reports per-case held-out outcomes
on the shared C/C++ benchmark, and Figure~\ref{fig:costcov} relates those
outcomes to repair cost for the same four systems. Section~\ref{app:related}
gives the conceptual comparison: Table~\ref{tab:iterate} summarizes what each
repair loop ranges over, what feedback it receives, and whether the repair
target can move.

\subsection{Empirical Re-Scoring Details}
\label{app:published-details}

The main text reports aggregate pairwise comparisons against \base,
\textsc{San2Patch}, LoopRepair, and \textsc{PatchAgent}. This subsection makes
the published-system comparison auditable at the case level. For each of the 41
C/C++ subjects, Table~\ref{tab:percase} reports whether each system blocks all,
some, or none of the certified paths in each oracle family, $\Cfix$
and $\Cres$. The table uses
the same symbolic grammar throughout: \full{} denotes full blocking,
\partialblk{} denotes partial blocking, \zeroblk{} denotes no blocking, \nopatch{}
denotes no reported patch, \na{} denotes no applicable certified suite or a
subject outside that system's benchmark, and \notmeas{} denotes a subject that
was not measured. Its subject set is the paper's, so
an unrepaired subject is a \nopatch{} row rather than an absent one --- including
the two \sys itself left unrepaired --- and its ``blocks every certified path'' totals
are the same counts that Table~\ref{tab:baselines} averages.

The comparator columns are sparse by construction: each published system is
scored only on the subjects its own benchmark contains --- 33 for
\textsc{San2Patch}, 40 for LoopRepair, and 17 for \textsc{PatchAgent} --- so
most of a comparator's column is not applicable at all, because the subject lies
outside that benchmark rather than because it carries no certified suite. One
\textsc{PatchAgent} cell is \notmeas{} rather than \nopatch{}: on coreutils
\texttt{gnu19784} its own functional gate is unusable, because the gate's
\texttt{test.sh} hard-codes an expected failure count that the \emph{unpatched}
tree does not reproduce in our environment, so its validator would reject every
candidate regardless of correctness. That subject was never attempted and
produced no \textsc{PatchAgent} patch, so it is excluded from the denominator
rather than scored as a repair failure --- which is why
Table~\ref{tab:baselines} reports $n = 16$ for \textsc{PatchAgent} under
$\Cfix$, and $n = 6$ under $\Cres$, the six of those subjects that carry a
certified residual suite.

\begin{figure}[t]
\centering
\includegraphics[width=\columnwidth]{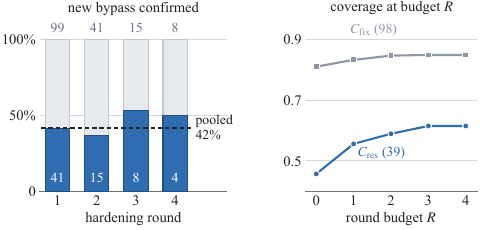}

\caption{Hardening-round dynamics. \textbf{Left:} execution-confirmed bypass
rate among runs entering each round; the pooled round-level rate is 42\%
(68/163), distinct from the 41\% run-level rate in
Section~\ref{sec:rq1}. \textbf{Right:} held-out coverage versus round budget,
with $R=0$ denoting PoV-only repair. Later-round cohorts are small; exact
values are in Table~\ref{tab:round-sweep}.}

\label{fig:attrition}
\end{figure}

Table~\ref{tab:percase} should be read as a shared-oracle re-scoring, not as a
reproduction of each paper's original evaluation. Prior systems used different
inputs and validation criteria: some were driven by crashing inputs or
sanitizer traces, while our main configuration starts from a normalized finding
and grows repair requirements through execution-confirmed bypasses. We therefore
score available patches or rerun artifacts under the same held-out suites to
measure what their original validation may not have tested.

\begin{figure}[t]
\centering
\includegraphics[width=\columnwidth]{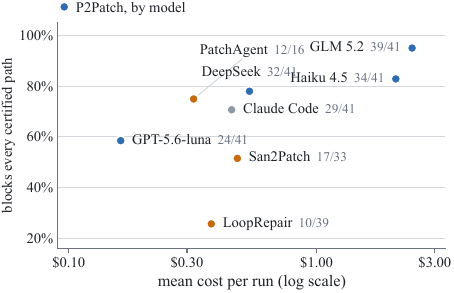}
\caption{Operational cost versus strict fix-closed coverage on the shared
C/C++ benchmark. Coverage is the fraction of subjects blocking every certified
fix-closed path under intention-to-treat scoring. \sys model points use all 41
subjects; published systems use their measurable shared-subject sets. Cost is
mean spend per actual run.}
\label{fig:costcov}
\end{figure}

\begin{table*}[!t]
\centering
\scriptsize
\renewcommand{\arraystretch}{0.94}
\caption{Per-case held-out outcomes on the 41 C/C++ subjects under
intention-to-treat scoring. \full{}, \partialblk{}, and \zeroblk{} denote full,
partial, and zero blocking; \nopatch{} denotes no produced patch, \na{} no
applicable suite or out-of-benchmark subject, and \notmeas{} an unmeasured
subject. A residual \zeroblk{} matches upstream behavior and is not a repair
failure. Shaded columns are \sys.}
\label{tab:percase}

\begin{tabular*}{\textwidth}{
@{\extracolsep{\fill}}
llc
cc
cc
cc
cc
>{\columncolor{p2blueLight}}c
>{\columncolor{p2blueLight}[\tabcolsep][0pt]}c
@{}
}
\toprule
 & & &
 \multicolumn{2}{c}{Claude Code} &
 \multicolumn{2}{c}{\textsc{San2Patch}} &
 \multicolumn{2}{c}{LoopRepair} &
 \multicolumn{2}{c}{\textsc{PatchAgent}} &
 \multicolumn{2}{>{\columncolor{p2blueLight}[\tabcolsep][0pt]}c}{\best{\sys}} \\
\cmidrule(lr){4-5}
\cmidrule(lr){6-7}
\cmidrule(lr){8-9}
\cmidrule(lr){10-11}
\cmidrule(lr){12-13}

Project & CVE & Type &
$\Cfix$ & $\Cres$ &
$\Cfix$ & $\Cres$ &
$\Cfix$ & $\Cres$ &
$\Cfix$ & $\Cres$ &
$\Cfix$ & $\Cres$ \\
\midrule

\multirow{2}{*}{libjpeg-turbo}
 & 2012-2806 & OOB-W
 & \full & \na
 & \full & \na
 & \full & \na
 & \full & \na
 & \full & \na \\

 & 2017-15232 & NPD
 & \partialblk & \na
 & \partialblk & \na
 & \nopatch & \nopatch
 & \partialblk & \na
 & \partialblk & \na \\

\midrule

\multirow{5}{*}{libxml2}
 & 2012-5134 & BOF
 & \zeroblk & \na
 & \full & \na
 & \zeroblk & \na
 & \full & \na
 & \full & \na \\

 & 2016-1833 & OOB-R
 & \zeroblk & \na
 & \na & \na
 & \zeroblk & \na
 & \na & \na
 & \nopatch & \na \\

 & 2016-1838 & OOB-R
 & \full & \na
 & \full & \na
 & \zeroblk & \na
 & \full & \na
 & \full & \na \\

 & 2016-1839 & OOB-R
 & \full & \na
 & \zeroblk & \na
 & \zeroblk & \na
 & \full & \na
 & \full & \na \\

 & 2017-5969 & NPD
 & \full & \full
 & \full & \zeroblk
 & \full & \zeroblk
 & \full & \zeroblk
 & \full & \full \\

\midrule

\multirow{1}{*}{potrace}
 & 2013-7437 & IOF
 & \partialblk & \zeroblk
 & \nopatch & \nopatch
 & \zeroblk & \zeroblk
 & \na & \na
 & \full & \zeroblk \\

\midrule

\multirow{16}{*}{libtiff}
 & 2016-10092 & BOF
 & \full & \na
 & \full & \na
 & \full & \na
 & \na & \na
 & \full & \na \\

 & 2016-10272 & BOF
 & \full & \na
 & \nopatch & \nopatch
 & \nopatch & \nopatch
 & \na & \na
 & \full & \na \\

 & 2016-3186 & BOF
 & \full & \na
 & \na & \na
 & \zeroblk & \na
 & \zeroblk & \na
 & \full & \na \\

 & 2016-5314 & OOB-W
 & \full & \zeroblk
 & \na & \na
 & \zeroblk & \zeroblk
 & \full & \zeroblk
 & \full & \full \\

 & 2016-5321 & BOF
 & \full & \full
 & \zeroblk & \full
 & \zeroblk & \full
 & \zeroblk & \full
 & \full & \full \\

 & 2016-9532 & OOB-R
 & \full & \na
 & \zeroblk & \na
 & \nopatch & \nopatch
 & \na & \na
 & \full & \na \\

 & 2017-5225 & BOF
 & \full & \zeroblk
 & \partialblk & \zeroblk
 & \partialblk & \zeroblk
 & \na & \na
 & \full & \full \\

 & 2017-7595 & DZ
 & \full & \na
 & \full & \na
 & \partialblk & \na
 & \full & \na
 & \full & \na \\

 & 2017-7599 & IV
 & \partialblk & \na
 & \partialblk & \na
 & \partialblk & \na
 & \na & \na
 & \full & \na \\

 & 2017-7600 & IV
 & \partialblk & \na
 & \partialblk & \na
 & \partialblk & \na
 & \na & \na
 & \partialblk & \na \\

 & 2017-7601 & IV
 & \full & \zeroblk
 & \partialblk & \zeroblk
 & \partialblk & \zeroblk
 & \partialblk & \zeroblk
 & \full & \full \\

 & 2018-18557 & OOB-W
 & \full & \na
 & \na & \na
 & \full & \na
 & \na & \na
 & \full & \na \\

 & 2022-4645 & OOB-R
 & \zeroblk & \na
 & \na & \na
 & \partialblk & \na
 & \na & \na
 & \full & \na \\

 & 2022-48281 & OOB-W
 & \full & \na
 & \na & \na
 & \full & \na
 & \na & \na
 & \full & \na \\

 & bz2611 & DZ
 & \full & \zeroblk
 & \full & \zeroblk
 & \full & \zeroblk
 & \full & \zeroblk
 & \nopatch & \nopatch \\

 & bz2633 & OOB-R
 & \full & \na
 & \nopatch & \nopatch
 & \zeroblk & \na
 & \full & \na
 & \full & \na \\

\midrule

\multirow{1}{*}{libarchive}
 & 2016-5844 & IOF
 & \full & \na
 & \full & \na
 & \zeroblk & \na
 & \na & \na
 & \full & \na \\

\midrule

\multirow{4}{*}{jasper}
 & 2016-8691 & DZ
 & \full & \na
 & \full & \na
 & \full & \na
 & \full & \na
 & \full & \na \\

 & 2016-9557 & IOF
 & \full & \na
 & \full & \na
 & \full & \na
 & \na & \na
 & \full & \na \\

 & 2020-27828 & OOB-W
 & \full & \na
 & \na & \na
 & \nopatch & \nopatch
 & \na & \na
 & \full & \na \\

 & 2021-3272 & OOB-R
 & \full & \na
 & \na & \na
 & \nopatch & \nopatch
 & \na & \na
 & \full & \na \\

\midrule

\multirow{3}{*}{libming}
 & 2016-9264 & BOF
 & \full & \zeroblk
 & \partialblk & \zeroblk
 & \full & \full
 & \na & \na
 & \full & \full \\

 & 2018-8806 & UAF
 & \partialblk & \na
 & \full & \na
 & \nopatch & \nopatch
 & \na & \na
 & \partialblk & \na \\

 & 2018-8964 & UAF
 & \partialblk & \full
 & \nopatch & \nopatch
 & \nopatch & \nopatch
 & \na & \na
 & \full & \full \\

\midrule

\multirow{4}{*}{binutils-gdb}
 & 2017-14745 & IOF
 & \partialblk & \na
 & \nopatch & \nopatch
 & \nopatch & \nopatch
 & \na & \na
 & \full & \na \\

 & 2017-15020 & OOB-R
 & \partialblk & \na
 & \partialblk & \na
 & \na & \na
 & \na & \na
 & \partialblk & \na \\

 & 2017-15025 & DZ
 & \full & \zeroblk
 & \full & \zeroblk
 & \nopatch & \nopatch
 & \full & \zeroblk
 & \full & \zeroblk \\

 & 2017-6965 & BOF
 & \full & \na
 & \full & \na
 & \nopatch & \nopatch
 & \na & \na
 & \full & \na \\

\midrule

\multirow{3}{*}{zziplib}
 & 2017-5974 & BOF
 & \full & \na
 & \zeroblk & \na
 & \zeroblk & \na
 & \na & \na
 & \full & \na \\

 & 2017-5975 & OOB-W
 & \partialblk & \zeroblk
 & \full & \na
 & \notmeas & \notmeas
 & \na & \na
 & \partialblk & \zeroblk \\

 & 2017-5976 & OOB-W
 & \full & \zeroblk
 & \full & \zeroblk
 & \zeroblk & \zeroblk
 & \na & \na
 & \full & \zeroblk \\

\midrule

\multirow{2}{*}{coreutils}
 & gnu19784 & OOB-R
 & \full & \na
 & \full & \na
 & \full & \na
 & \notmeas & \notmeas
 & \full & \na \\

 & gnu25023 & OOB-R
 & \full & \na
 & \full & \na
 & \zeroblk & \na
 & \full & \na
 & \full & \na \\

\midrule

\multicolumn{3}{@{}l}{\textbf{Blocks every certified path}}
 & \multicolumn{2}{c}{29}
 & \multicolumn{2}{c}{17}
 & \multicolumn{2}{c}{10}
 & \multicolumn{2}{c}{12}
 & \multicolumn{2}{>{\columncolor{p2blueLight}[\tabcolsep][0pt]}c}{\best{34}} \\

\multicolumn{3}{@{}l}{\textbf{Blocks some, not all}}
 & \multicolumn{2}{c}{9}
 & \multicolumn{2}{c}{7}
 & \multicolumn{2}{c}{6}
 & \multicolumn{2}{c}{2}
 & \multicolumn{2}{>{\columncolor{p2blueLight}[\tabcolsep][0pt]}c}{5} \\

\multicolumn{3}{@{}l}{\textbf{Blocks none}}
 & \multicolumn{2}{c}{3}
 & \multicolumn{2}{c}{4}
 & \multicolumn{2}{c}{13}
 & \multicolumn{2}{c}{2}
 & \multicolumn{2}{>{\columncolor{p2blueLight}[\tabcolsep][0pt]}c}{0} \\

\multicolumn{3}{@{}l}{\textbf{No patch produced}}
 & \multicolumn{2}{c}{0}
 & \multicolumn{2}{c}{5}
 & \multicolumn{2}{c}{10}
 & \multicolumn{2}{c}{0}
 & \multicolumn{2}{>{\columncolor{p2blueLight}[\tabcolsep][0pt]}c}{2} \\

\multicolumn{3}{@{}l}{\textbf{Not measured}}
 & \multicolumn{2}{c}{0}
 & \multicolumn{2}{c}{0}
 & \multicolumn{2}{c}{1}
 & \multicolumn{2}{c}{1}
 & \multicolumn{2}{>{\columncolor{p2blueLight}[\tabcolsep][0pt]}c}{0} \\

\bottomrule
\end{tabular*}
\end{table*}

\begin{figure*}[!t]
\centering
\includegraphics[width=\textwidth]{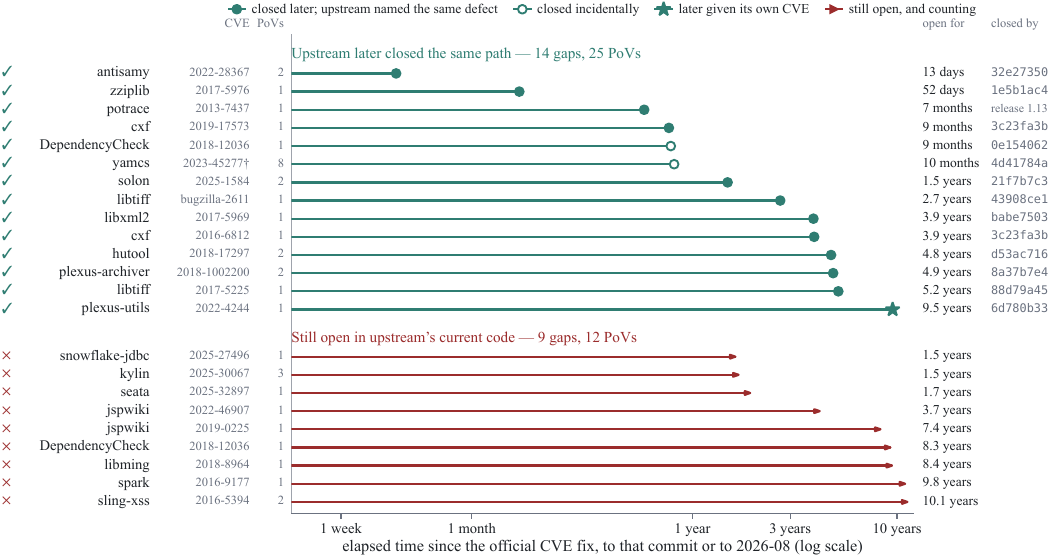}
\caption{Residual gaps against upstream history. Each row is one gap the
official CVE fix left open, and its length is how long that gap survived in
upstream's own code: up to the commit that later closed the same path
($\checkmark$, green), or up to August 2026 with the path still there ($\times$,
red). A hollow marker is a commit that closed the path without recording that a
defect existed. The \povs\ column counts the certified residual \povs driving
that one gap, and the last column is the upstream commit that closed it. $^{\dagger}$~the same
gap is certified under CVE-2023-45278 as well; those two suites share one
official fix and one closing commit. potrace has no public repository, so its
row is dated release to release.}
\label{fig:timeline}
\end{figure*}

\begin{table*}[tp]
\centering
\footnotesize
\caption{Recent systems all iterate. What differs is what the loop ranges over,
what feedback it receives, and whether the repair target can move.}
\label{tab:iterate}
\setlength{\tabcolsep}{5pt}
\begin{tabularx}{\textwidth}{@{}lLLc@{}}
\toprule
\headrow
System & Loop ranges over & Feedback & Objective moves? \\
\midrule
\textsc{San2Patch}     & reasoning paths (ToT) & compiler, PoC result & no \\
\textsc{PatchAgent}    & agent actions & test failure, \emph{rejected patches} & no \\
LoopRepair             & candidate patches & ranked failing patch & no \\
\midrule
\textbf{\sys}          & \textbf{the objective} & \textbf{a new working exploit} & \textbf{yes} \\
\bottomrule
\end{tabularx}
\end{table*}

Figure~\ref{fig:costcov} provides the corresponding cost--coverage diagnostic
on the same shared C/C++ benchmark. It uses a strict coverage reading: the
fraction of subjects whose patch blocks every certified fix-closed path,
counting a subject the system left unrepaired against it. This view separates two
questions that are otherwise easy to conflate: whether a method can ship a patch
cheaply, and whether that patch covers all certified paths under the independent
oracle suite.

The cost view is diagnostic rather than primary. Our main claim is not that
counterexample-guided repair is cheaper than one-shot repair; the protocol runs
additional exploitation and correction stages by design. Instead,
Figure~\ref{fig:costcov} shows the coverage obtained for that extra work under
a common oracle. The comparison is therefore end-to-end: it reflects differences
in input modality, repair protocol, implementation maturity, and validation
policy, not only the adversarial hardening loop.

\subsection{Conceptual Comparison}
\label{app:related}

Table~\ref{tab:iterate} summarizes the conceptual distinction between \sys and
recent iterative repair systems. The key question is not whether a system
iterates---all recent agentic systems do---but what the loop ranges over and
whether the target can change after a patch is accepted. Prior systems iterate
over reasoning paths, agent actions, candidate patches, or failed patches while
the success target remains fixed. \sys instead changes the repair objective:
each execution-confirmed bypass of the current patch becomes a new repair
requirement.

This distinction also explains why the empirical and conceptual comparisons are
both needed. The empirical re-scoring asks how prior patches behave under our
held-out certified oracles. The conceptual comparison explains why \sys may
produce different patches in the first place: the target is not fixed to the
initial alert, trace, crash, or \pov, but can expand whenever the current
patch is execution-falsified.

\section{Per-Subject Outcomes}
\label{app:persubject}

Tables~\ref{tab:persubject} and~\ref{tab:persubjectc} report all 101 subjects
individually (60 Java, 41 C/C++), so the aggregates of
Section~\ref{sec:eval} can be checked against the underlying distribution.
The three subjects \sys left unrepaired are marked \nopatch{} and carry zero
coverage in every aggregate, so the two tables sum to
Table~\ref{tab:coverage}.

\begin{table*}[tp]
\centering
\footnotesize
\caption{Per-subject held-out coverage for 60 Java subjects. $|\Sfix|$ is the
fix-closed suite size and $b=|E|-1$ the number of appended execution-confirmed
bypasses. ``base'' is \base and ``ours'' is \sys; bold marks improvement,
$\downarrow$ regression, \na{} no residual suite, and \nopatch{} no accepted
\sys patch. Residual suite sizes are in Table~\ref{tab:benchmark}.}
\label{tab:persubject}
\scriptsize
\setlength{\tabcolsep}{3.5pt}
\renewcommand{\arraystretch}{1.05}
\begin{tabular}[t]{@{}llrrrrrr@{}}
\toprule
\headrow
CVE & Project & $|\Sfix|$ & \multicolumn{2}{c}{$\Cfix$} & \multicolumn{2}{c}{$\Cres$} & $b$ \\
\cmidrule(lr){4-5}\cmidrule(lr){6-7}
 & & & base & ours & base & ours & \\
\midrule
2014-7816 & undertow & 3 & 1.00 & 1.00 & \na & \na & 0 \\
\rowcolor{p2gray}2016-5394 & sling-xss & 3 & 1.00 & 1.00 & 0.00 & 0.00 & 0 \\
2016-6812 & cxf & 2 & 0.50 & 0.50 & 1.00 & 1.00 & 4 \\
\rowcolor{p2gray}2016-9177 & spark & 2 & 0.50 & \textbf{1.00} & 1.00 & 1.00 & 0 \\
2017-1000487 & plexus-utils & 8 & 0.50 & \textbf{1.00} & \na & \na & 1 \\
\rowcolor{p2gray}2018-1000129 & jolokia & 6 & 0.67 & 0.67 & \na & \na & 0 \\
2018-1002200 & plexus-archiver & 4 & 1.00 & 1.00 & 1.00 & 1.00 & 1 \\
\rowcolor{p2gray}2018-1002201 & zt-zip & 4 & 1.00 & 1.00 & \na & \na & 1 \\
2018-1002202 & zip4j & 3 & 0.67 & \textbf{1.00} & 0.50 & \textbf{1.00} & 0 \\
\rowcolor{p2gray}2018-1047 & wildfly & 3 & 1.00 & 1.00 & 1.00 & 1.00 & 0 \\
2018-12036 & DependencyCheck & 6 & 0.33 & 0.33 & 0.50 & 0.50 & 0 \\
\rowcolor{p2gray}2018-12542 & vertx-web & 3 & 1.00 & 1.00 & \na & \na & 1 \\
2018-1260 & spring-security-oauth & 2 & 1.00 & 1.00 & \na & \na & 0 \\
\rowcolor{p2gray}2018-17297 & hutool & 5 & 0.80 & 0.20$\downarrow$ & 1.00 & 0.50$\downarrow$ & 0 \\
2018-9159 & spark & 5 & 0.00 & \textbf{1.00} & \na & \na & 2 \\
\rowcolor{p2gray}2019-0207 & tapestry-5 & 2 & 1.00 & 1.00 & \na & \na & 0 \\
2019-0225 & jspwiki & 1 & 0.00 & \textbf{1.00} & 0.00 & 0.00 & 0 \\
\rowcolor{p2gray}2019-10077 & jspwiki & 3 & 1.00 & 1.00 & \na & \na & 1 \\
2019-17573 & cxf & 6 & 1.00 & 1.00 & 0.00 & 0.00 & 1 \\
\rowcolor{p2gray}2020-2261 & perfecto-plugin & 1 & 1.00 & 1.00 & 1.00 & 1.00 & 1 \\
2020-35460 & mpxj & 4 & 1.00 & 1.00 & \na & \na & 0 \\
\rowcolor{p2gray}2020-8570 & kubernetes-java & 2 & 1.00 & 1.00 & \na & \na & 0 \\
2021-29425 & commons-io & 4 & 0.00 & 0.00 & \na & \na & 1 \\
\rowcolor{p2gray}2021-41269 & cron-utils & 2 & 1.00 & 1.00 & \na & \na & 1 \\
2022-22932 & karaf & 3 & 0.67 & 0.67 & \na & \na & 0 \\
\rowcolor{p2gray}2022-22947 & spring-cloud-gateway & 3 & 1.00 & 1.00 & \na & \na & 0 \\
2022-23082 & curekit & 2 & 1.00 & 1.00 & \na & \na & 1 \\
\rowcolor{p2gray}2022-25842 & one-java-agent & 3 & 1.00 & 1.00 & \na & \na & 0 \\
2022-26049 & goomph & 2 & 1.00 & 1.00 & \na & \na & 0 \\
\rowcolor{p2gray}2022-26884 & dolphinscheduler & 7 & 1.00 & \nopatch & 1.00 & \nopatch & 0 \\
\bottomrule
\end{tabular}
\hspace{9mm}
\begin{tabular}[t]{@{}llrrrrrr@{}}
\toprule
\headrow
CVE & Project & $|\Sfix|$ & \multicolumn{2}{c}{$\Cfix$} & \multicolumn{2}{c}{$\Cres$} & $b$ \\
\cmidrule(lr){4-5}\cmidrule(lr){6-7}
 & & & base & ours & base & ours & \\
\midrule
2022-28367 & antisamy & 2 & 0.00 & \textbf{1.00} & 0.00 & \textbf{1.00} & 1 \\
\rowcolor{p2gray}2022-32287 & uima-uimaj & 3 & 1.00 & 1.00 & \na & \na & 0 \\
2022-4244 & plexus-utils & 3 & 1.00 & 1.00 & 1.00 & 1.00 & 0 \\
\rowcolor{p2gray}2022-42889 & commons-text & 12 & 0.33 & 0.33 & \na & \na & 2 \\
2022-44262 & ff4j & 10 & 0.70 & \textbf{1.00} & \na & \na & 3 \\
\rowcolor{p2gray}2022-46166 & spring-boot-admin & 14 & 0.00 & \textbf{0.50} & \na & \na & 2 \\
2022-46907 & jspwiki & 7 & 0.29 & 0.29 & 0.00 & 0.00 & 0 \\
\rowcolor{p2gray}2022-4963 & spring-module-core & 3 & 1.00 & 1.00 & \na & \na & 1 \\
2023-24422 & script-security-plugin & 3 & 0.00 & 0.00 & \na & \na & 0 \\
\rowcolor{p2gray}2023-37460 & plexus-archiver & 2 & 1.00 & 1.00 & \na & \na & 0 \\
2023-37908 & xwiki-rendering & 3 & 0.00 & 0.00 & 0.00 & 0.00 & 4 \\
\rowcolor{p2gray}2023-45277 & yamcs & 6 & 1.00 & 1.00 & 0.00 & \textbf{1.00} & 2 \\
2023-45278 & yamcs & 5 & 1.00 & 1.00 & 0.00 & \textbf{1.00} & 3 \\
\rowcolor{p2gray}2025-0851 & djl & 2 & 0.00 & 0.00 & 0.00 & 0.00 & 0 \\
2025-1391 & keycloak & 3 & 0.33 & 0.33 & \na & \na & 1 \\
\rowcolor{p2gray}2025-1584 & solon & 1 & 1.00 & 1.00 & 0.50 & \textbf{1.00} & 3 \\
2025-23026 & jte & 4 & 0.25 & \textbf{0.50} & 0.00 & \textbf{1.00} & 3 \\
\rowcolor{p2gray}2025-26795 & iotdb & 3 & 1.00 & 1.00 & \na & \na & 0 \\
2025-26864 & iotdb & 1 & 1.00 & 1.00 & \na & \na & 1 \\
\rowcolor{p2gray}2025-27496 & snowflake-jdbc & 1 & 1.00 & 1.00 & 0.00 & 0.00 & 4 \\
2025-27508 & emissary & 4 & 0.75 & 0.75 & 1.00 & 1.00 & 2 \\
\rowcolor{p2gray}2025-30067 & kylin & 2 & 0.00 & \textbf{0.50} & 0.33 & \textbf{1.00} & 1 \\
2025-32897 & incubator-seata & 2 & 1.00 & 1.00 & 1.00 & 1.00 & 0 \\
\rowcolor{p2gray}2025-32960 & restapi & 2 & 1.00 & 1.00 & 1.00 & 1.00 & 0 \\
2025-32961 & jpawebapi & 1 & 1.00 & 1.00 & 1.00 & 1.00 & 4 \\
\rowcolor{p2gray}2025-46096 & solon & 3 & 1.00 & 1.00 & \na & \na & 0 \\
2025-4641 & webdrivermanager & 1 & 1.00 & 1.00 & \na & \na & 1 \\
\rowcolor{p2gray}2025-47771 & powsybl-core & 1 & 1.00 & 1.00 & \na & \na & 0 \\
2025-52888 & allure2 & 3 & 1.00 & 1.00 & \na & \na & 0 \\
\rowcolor{p2gray}2025-53689 & jackrabbit & 2 & 0.50 & 0.50 & 0.00 & 0.00 & 0 \\
\bottomrule
\end{tabular}
\end{table*}

\begin{table*}[tp]
\centering
\footnotesize
\caption{Per-subject held-out coverage for 41 C/C++ subjects; notation follows
Table~\ref{tab:persubject}.}
\label{tab:persubjectc}

\scriptsize
\renewcommand{\arraystretch}{1.05}

\begin{tabular*}{0.475\textwidth}{@{\extracolsep{\fill}}llrrrrrr@{}}
\toprule
\headrow
CVE & Project & $|\Sfix|$ & \multicolumn{2}{c}{$\Cfix$} &
\multicolumn{2}{c}{$\Cres$} & $b$ \\
\cmidrule(lr){4-5}\cmidrule(lr){6-7}
 & & & base & ours & base & ours & \\
\midrule
2012-2806 & libjpeg-turbo & 1 & 1.00 & 1.00 & \na & \na & 0 \\
\rowcolor{p2gray}2012-5134 & libxml2 & 1 & 0.00 & \textbf{1.00} & \na & \na & 0 \\
2013-7437 & potrace & 6 & 0.83 & \textbf{1.00} & 0.00 & 0.00 & 1 \\
\rowcolor{p2gray}2016-10092 & libtiff & 1 & 1.00 & 1.00 & \na & \na & 0 \\
2016-10272 & libtiff & 1 & 1.00 & 1.00 & \na & \na & 0 \\
\rowcolor{p2gray}2016-1833 & libxml2 & 4 & 0.00 & \nopatch & \na & \na & 0 \\
2016-1838 & libxml2 & 1 & 1.00 & 1.00 & \na & \na & 1 \\
\rowcolor{p2gray}2016-1839 & libxml2 & 4 & 1.00 & 1.00 & \na & \na & 0 \\
2016-3186 & libtiff & 2 & 1.00 & 1.00 & \na & \na & 0 \\
\rowcolor{p2gray}2016-5314 & libtiff & 2 & 1.00 & 1.00 & 0.00 & \textbf{1.00} & 0 \\
2016-5321 & libtiff & 2 & 1.00 & 1.00 & 1.00 & 1.00 & 0 \\
\rowcolor{p2gray}2016-5844 & libarchive & 2 & 1.00 & 1.00 & \na & \na & 0 \\
2016-8691 & jasper & 2 & 1.00 & 1.00 & \na & \na & 0 \\
\rowcolor{p2gray}2016-9264 & libming & 3 & 1.00 & 1.00 & 0.00 & \textbf{1.00} & 1 \\
2016-9532 & libtiff & 1 & 1.00 & 1.00 & \na & \na & 0 \\
\rowcolor{p2gray}2016-9557 & jasper & 1 & 1.00 & 1.00 & \na & \na & 0 \\
2017-14745 & binutils-gdb & 2 & 0.50 & \textbf{1.00} & \na & \na & 0 \\
\rowcolor{p2gray}2017-15020 & binutils-gdb & 3 & 0.33 & \textbf{0.67} & \na & \na & 1 \\
2017-15025 & binutils-gdb & 1 & 1.00 & 1.00 & 0.00 & 0.00 & 0 \\
\rowcolor{p2gray}2017-15232 & libjpeg-turbo & 7 & 0.86 & 0.86 & \na & \na & 1 \\
2017-5225 & libtiff & 2 & 1.00 & 1.00 & 0.00 & \textbf{1.00} & 0 \\
\bottomrule
\end{tabular*}
\hspace{5mm}
\begin{tabular*}{0.475\textwidth}{@{\extracolsep{\fill}}llrrrrrr@{}}
\toprule
\headrow
CVE & Project & $|\Sfix|$ & \multicolumn{2}{c}{$\Cfix$} &
\multicolumn{2}{c}{$\Cres$} & $b$ \\
\cmidrule(lr){4-5}\cmidrule(lr){6-7}
 & & & base & ours & base & ours & \\
\midrule
\rowcolor{p2gray}2017-5969 & libxml2 & 1 & 1.00 & 1.00 & 1.00 & 1.00 & 1 \\
2017-5974 & zziplib & 1 & 1.00 & 1.00 & \na & \na & 0 \\
\rowcolor{p2gray}2017-5975 & zziplib & 2 & 0.50 & 0.50 & 0.00 & 0.00 & 0 \\
2017-5976 & zziplib & 2 & 1.00 & 1.00 & 0.00 & 0.00 & 0 \\
\rowcolor{p2gray}2017-6965 & binutils-gdb & 4 & 1.00 & 1.00 & \na & \na & 0 \\
2017-7595 & libtiff & 3 & 1.00 & 1.00 & \na & \na & 0 \\
\rowcolor{p2gray}2017-7599 & libtiff & 8 & 0.88 & \textbf{1.00} & \na & \na & 1 \\
2017-7600 & libtiff & 8 & 0.75 & 0.75 & \na & \na & 1 \\
\rowcolor{p2gray}2017-7601 & libtiff & 5 & 1.00 & 1.00 & 0.00 & \textbf{1.00} & 2 \\
2018-18557 & libtiff & 1 & 1.00 & 1.00 & \na & \na & 0 \\
\rowcolor{p2gray}2018-8806 & libming & 6 & 0.67 & 0.67 & \na & \na & 0 \\
2018-8964 & libming & 5 & 0.80 & \textbf{1.00} & 1.00 & 1.00 & 1 \\
\rowcolor{p2gray}2020-27828 & jasper & 2 & 1.00 & 1.00 & \na & \na & 0 \\
2021-3272 & jasper & 2 & 1.00 & 1.00 & \na & \na & 0 \\
\rowcolor{p2gray}2022-4645 & libtiff & 3 & 0.00 & \textbf{1.00} & \na & \na & 2 \\
2022-48281 & libtiff & 1 & 1.00 & 1.00 & \na & \na & 0 \\
\rowcolor{p2gray}bz2611 & libtiff & 2 & 1.00 & \nopatch & 0.00 & \nopatch & 0 \\
bz2633 & libtiff & 1 & 1.00 & 1.00 & \na & \na & 1 \\
\rowcolor{p2gray}gnu19784 & coreutils & 3 & 1.00 & 1.00 & \na & \na & 0 \\
gnu25023 & coreutils & 2 & 1.00 & 1.00 & \na & \na & 0 \\
\bottomrule
\end{tabular*}

\end{table*}

\section{Residual Gaps and Upstream History}
\label{app:upstream}

A residual \pov certifies that a path remains exploitable after the official
fix, which makes it a valid measuring instrument but does not by itself
establish that the gap mattered. We therefore manually reviewed every certified
residual gap against upstream's current code and history.
Figure~\ref{fig:timeline} plots the outcome, restricted to the gaps whose
disposition we could settle with confidence and whose claim falls inside the
CVE's own scope --- 37 \povs over 23 gaps in 19 projects; gaps that no released
tree ever carried, and those we could not settle, are left out.

For 14 of those gaps --- 25 \povs\ --- upstream later closed exactly the path
our \pov drives. The interval from the official fix to that commit is a
detection lead: our suites flag at $t = 0$ what upstream removed a
median of just over two years later, and in one case nine and a half years
later. Twelve of the fourteen closing commits, or the issues they close, name
the same defect, so upstream itself agrees the path was a defect rather than a
refactor. One went further: the plexus-utils gap was independently rediscovered
and assigned CVE-2025-67030, nine and a half years after the fix that left it
open.

The remaining 9 gaps --- 12 \povs\ --- are still present at upstream's head,
between a year and a half and ten years after the fix meant to close them, one
in a maintained and widely deployed dependency-scanning tool. All 9 have been
reported to their projects (Section~\ref{sec:ethics}).

\enlargethispage{2\baselineskip}
Two caveats apply. Open at head is not the same as exploitable: several of
these gaps drive a library API or command-line tool rather than a product entry
point, and reachability from attacker-controlled input must be established case
by case before any one is a finding. And intervals are measured commit to
commit, not as a user experiences them; release-to-release exposure can differ
sharply, as the antisamy row shows, where 13 days between commits still let two
releases ship carrying the gap.

\end{document}